\documentclass[10pt, leqno]{amsart}
 
\usepackage{amsfonts}
\usepackage{verbatim,amssymb,amsmath}
\numberwithin{equation}{section}
\usepackage{amscd}

\newtheorem{thm}{Theorem}[section]
\newtheorem{lem}[thm]{Lemma}

\newtheorem{defn}{Definition}[section]
\newtheorem{claim}{Claim}[section]
\newtheorem{remark}{Remark}[section]

\newtheorem{problem}{Problem}[section]

\begin{document}

\title[Does there exist the applicability limit of PDE] 
{{Does there exist the applicability limit}\\[1ex] 
{of PDE to describe physical phenomena?}\\[1ex] 
{\small --- A personal survey of}\\
{\small Quantization, QED, Turbulence --- }}

\author{Atsushi Inoue} 
\thanks{Author: alias atlom=a tiny little old mathematician, or suman=superman$-$``per", ``per'' pronounces  ``pa-"  in Japanese, 
means ``bone head" or ``stupid", therefore deleted and finally ``suman'' stands for ``sorry" in Japanese, curious!}
\dedicatory{Dedication to Masuda san for his warm consideration}
\date {\today, based on a talk on 19 March  2023}
\address{Professor Emeritus, Department of Mathematics, Tokyo Institute of Technology}
\subjclass{MSC-class, 35,42,46,76}
\email{inoue@math.titech.ac.jp}

\def\where{\quad\text{where}\quad}
\def\when{\quad\text{when}\quad}
\def\with{\quad\text{with}\quad}
\def\for{\quad\text{for}\quad}
\def\forany{\quad\text{for any}\quad}
\def\foreach{\quad\text{for each}\quad}
\def\et{\;\mbox{and}\;}
\def\bysame{--------}
\def\supp{\rm{supp}\,}
\def\dist{\,\text{dist}\,}
\def\euc{{\mathbb{R}}}
\def\fR{{\mathfrak{R}}}
\def\fC{{\mathfrak{C}}}
\def\rev{{\fR}_{\rm{ev}}}
\def\rod{{\fR}_{\rm{od}}}

\def\unbt{\underline{t}}
\def\unbx{\underline{x}}
\def\unbq{\underline{q}}
\def\unbxi{\underline{\xi}}
\def\unbp{\underline{p}}
\def\barx{\bar{x}}
\def\barq{\bar{q}}
\def\barxi{\bar{\xi}}
\def\unbeta{\underline{\eta}}
\def\unbtheta{\underline{\theta}}
\def\unbpi{\underline{\pi}}
\def\bartheta{\bar{\theta}}

\def\unbu{\underline{u}}

\def\tr{\operatorname{tr\,}}
\def\str{\operatorname{str\,}}

\def\spin{\kbar}  
\def\clifpin{{\mathchar'26\mkern-9mu {\lambda}}}
\def\kbar{{\mathchar'26\mkern-9mu k}}

\def\sdet{\operatorname{sdet}}

\def\frachi{\frac{\hbar}{i}}
\def\dt{\frac{d}{dt}}

\def\pdt{\frac{\partial}{\partial t}}
\def\pdx{\frac{\partial}{\partial x}}
\def\pdq{\frac{\partial}{\partial q}}
\def\pdxj{\frac{\partial}{\partial x_j}}

\def\pdq{\frac{\partial}{\partial {q}}}
\def\pdx{\frac{\partial}{\partial {x}}}

\def\pdt{\frac{\partial}{\partial {t}}}
\def\pdqj{\frac{\partial}{\partial {q_j}}}
\def\dx{\frac{d}{dx}}
\def\dt{\frac{d}{dt}}
\def\ds{\frac{d}{ds}}

\maketitle

\baselineskip=16pt 

\begin{abstract}
{What does it mean to study PDE(=Partial Differential Equation)? How and what to do ``to claim proudly that I'm studying a certain PDE''?
Newton mechanic uses mainly ODE(=Ordinary Differential Equation) and describes nicely movements of Sun, Moon and Earth etc.
Now, so-called quantum phenomenum is described by, say Schr\"odinger equation, PDE which explains
both wave and particle characters after quantization of ODE.
The coupled Maxwell-Dirac equation is also ``quantized'' and QED(=Quantum Electro-Dynamics) theory is invented by physicists. Though it is said this QED gives very good coincidence between theoretical\footnotemark
 and experimental observed quantities, but what is the equation corresponding to QED?
Or, is it possible to describe QED by ``equation" in naive sense?}
\end{abstract}
\footnotetext{Seemingly, \underline{physicists try to obtain theoretical values by perturbation series w.r.t. coupling constant but not necessarily} 
 \underline{``solving'' equation mathematically}. After H. Lewy's example, \underline{we can't believe so naively physicist's claim that there should} \underline{exist a solution of that PDE} even if physicist insists that equation derived properly from physical phenomena.}

\tableofcontents

\section{Introduction with a brief personal history of atlom}
Though I attended with the TV course of collegial physics\footnote{A part of NBC's educational program ``Continental Classroom'' in1958--1963, which is broadcasted from NHK in 1959--1961 as ``Physics in Nuclear Age'' with explanation in Japanese} in my high school time, as I felt the score of  Shingaku-Furiwake\footnote{Though depending on each University,  after 18months of general education entering  there, each student should decide to proceed in which department} was insufficient to go to physics\footnote{The main reason of shortage of score is caused by inmatured recognition of the level of English reading in general education which is seemingly at first glance lower than that of  high school, therefore I cut corners in that course! Be careful young people.}.
 I had no alternative but to proceed to applied mathematics because for me at that time, two branches seem to treat analogous objects. On the other hand, 
my career as a mathematician has been started rather accidentally. Despite I couldn't imagine I might be an ``ordinary'' salaried man\footnote{Seemingly, I'm not good at temporarily going along with ``respectable person" as an amenable member of organization.}, but fortunately perhaps, just after finished my master course, I was proposed a research assistant post of department of mathematics. 
Such posts made increased suddenly because of ``Science and Engineering Faculty Expansion Plan"  by the Japanese government which began after the Sputnik shock suffered by USA.  By the way, as is said  ``The teacher of the university is not resigned as once if he does it" or ``Beggars and monks can't quit after three days", I couldn't only resign from research assistant post but also had an opportunity to study as a boursier du gouvernement fran\c{c}ais. 

But after several years at around age thirties, I was offered a new position. At that time, continuously getting salary as a researcher, I should do mathematics more seriously, so I felt, and I started to study not only linear PDE but also non-linear PDE. Surely, H\"ormander's works on linear PDE are overwhelming at that time.

So to restart with a little mathematician, I questioned naively why the Navier-Stokes equation is so famous and I wondered if this equation is genuinely good enough or worth studying? 
If it is so good as equation, it should be invariant under change of variables? At that time, the initial and boundary value problem in the time dependent domain for the Navier-Stokes equation is studied by Fujita and Sauer \cite{FS70} 
by penalty method. Since I couldn't appreciate their method fully\footnote{Whether their penalty method works when $\cup_{0\le t\le T}\Omega(t)\times \{t\}\subset \euc^{d+1}$ when $\Omega(t)$ and $\Omega(t')$ is not necessarily diffeomorphic each other but with the same volume?}, I find the change of variable formula for vector field or differential 1-form with the help of Wakimoto \cite{IW77}
under the condition that ``gold fishes in the bowl don't allow even kissing\footnote{That is, assuming that the volume of $\Omega(t)$ is constant in $t$ and for any $t, t'$, $\Omega(t)$ and $\Omega(t')$ are diffeomorphic each other}''!
Not only this, even turbulent phenomena are believed to be governed by this same equation, why so?
Though this equation is derived from conservations of momentum and mass by observing ``laminar flow", why so easily believed the totally different looking phenomena, ``turbulent flows", are also governed by the same equation?
Not only this, the viscosity occurs after or before viewing the discrete molecular structure of water as a continuum\footnote{A physicist says naively that viscosity comes from particle structure of water, they don't bother when they,  to treat fluids, started to regard particles as if continuum. To consider this, I wrote \cite{IF79} with the aid of Funaki.}?
Here it is also appropriate quoting Hopf \cite{hop52} saying:
\begin{quotation}
Statistical mechanics constructs certain ``relevant" phase distributions which characterize the ``typical" phase motions and which must be used for all statistical predictions. $\cdots\cdots$
$\cdots$, in statistical hydromechanics -- the theory of highly turbulent fluid flow -- the small scale on which the ``fluid elements" interact seems to be of decisive importance, The relevant distributions based on this scale -- the hypothetical Kolmogoroff distributions -- must be mathematically very different from canonical distribution. So far all attempts to determine the relevant hydrodynamical phase distributions, at least to a sufficient degree of approximation, have met with considerable mathematical difficulties.
\end{quotation}

 From these consideration, I wonder whether ``is there applicability limit for describing physical phenomena by PDE''? and what is the limit of regarding huge number of water molecules as continuum? 

As is well-known, now called ``quantum phenomena'' are not so well described directly by ODE(Newton Mechanics), therefore we need to use PDE, Schr\"odinger equation(in some sense, a quantized version of Newton Mechanics).
Therefore, at least logically, there will exist some phenomena assumed to be not well describable using PDE.

For such phenomena, Gelfand \cite{gel54} not only proposed to use FDE(=Functional Derivative Equation) for understanding turbulence and QED, but also questioned whether our existed mathematics tools sufficient to describe these phenomena\footnote{Someone wrote that, ``if there doesn't exist Riemannian geometry or invariant theory, theory of relativity, if there doesn't exist boundary value theory, wave mechanics, and if there doesn't exist matrix theory, quantum mechanics may not be invented? New physical theory is stimulated by precedent mathematical method, or new mathematical method is invented from newly recognized phyisical phenomena''. Analogous mentioning  may be found in Sawyer \cite{saw55}}? 
I completely empathize with his opinion, and I feel we need to develop theory of FDE\footnote{But in general, we can't give meaning higher order functional derivatives at each point. Moreover, in infinite dimensional  topological spaces, there doesn't exist Lebesgue-like measure which permit integration by parts, Smolyanov and  Fomin \cite{SF76}. 
This means any trial to extend ``A study of PDE by functional analytic method''  seems breakdown from the outset. In spite of this, physicists, using tools not yet mathematically justified, get certain theoretical values and experimental values with complicated and expensive experiments, and astonishingly these values coincide many digits. This fact seems to imply something exist which is not yet appreciated mathematically.}. 

Apart from above, to treat initial value problem for linear hyperbolic systems of PDE about 50's--90's, there is a trend to diagonalize that system with posing conditions on properties of characteristic roots, as mathematical technique.
But I feel strange why we need to diagonalize system of PDE.
I feel curious to such treatise because not only until when such efforts continue but also there exists not diagonalizable system of  PDE.
From my point of view, the necessity of diagonalization is a desire to apply existing theory of pseudo-differential equations or more precisely, standard symbol calculus is only confined to scalar case.
Therefore, we need to treat matrix structure as it is, this is one motivation to construct superanalysis(=analysis on superspace $\mathfrak{R}^{m|n}$ not on $\euc^m$. Here $m$ or $n$ are symbolically used as something like ``dimension").

Even almost 30 years passed after I started to concern with FDE or superanalysis, these subjects never belong to the main stream of researches in mathematical society, at least in Japan.
Therefore, it seems natural, at that time, judges of KAKENHI(=Grants-in-Aid for Scientific Research in Japan) disregarded my proposal concerning FDE and superanalysis almost completely, so I think. But any way, none of them, even personally, asked me what are them? Finally, I proposed to ``make a room to accept disagreement to reviewers' evaluation for KAKENHI proposal"
\footnote{examiners' names are announced after one year later, contrary to ''Strike while the iron is hot''}
which is also neglected completely\footnote{Recall the following anecdote: The miraculous ability of Ramanujan, like decomposing taxi number as $1729=12^3+1^3=10^3+9^3$, was neglected as absurd by many professors except Hardy's intuitive recognition}. Under these  situations, K\^yuya Masuda supported me behind the scenes.

In this note, mathematics which is disregarded by judges of KAKENHI, 
are mentioned, because without doing so, these trials are in vain with my physical lifespan.
 
Since originally, the descriptive ability of PDE and its reproducibility of physical phenomena are main concern, it seems natural we wonder the difference between the derivation of classical field equation and quantum field mechanics.

In any way, it seems worth mentioning the following description which is given
in Functional Methods, Chapter 9 of Itzykson and Zuber \cite{IZ79}:
\begin{quotation}``The path integral formalism of Feynman and Kac provides a unified view of quantum mechanics, field theory, and statistical models. Starting from the case of finitely many degrees of freedom it is generalized to include fermion systems and then extended to infinite systems. The steepest-descent method of integration exhibits the close relationship with classical mechanics and allows us to recover ordinary perturbation theory.''
\end{quotation}

\section{Quantization and PIM(=Path Integral Method)}
\subsection{The beginning of PIM}
Following explanation is due to Feynman and Hibbs \cite{FH65}
but I cited here from Albeverio and Hoegh-Krohn \cite{AH76}.

Let consider the representation formula for the solution of Schr\"odinger equation on $\euc^d$
$$
i\hbar\frac{\partial}{\partial t}\psi(q,t)=-\frac{\hbar^2}{2m}\Delta \psi(q,t)+V(q)\psi(q,t),
$$
 with the initial data $\psi(q,0)=\underline{\psi}(q)$. Decomposing
$$
H=H_0+V \with H_0=-\frac{\hbar^2}{2m}\Delta,
$$
and assuming that $H$ is selfadjoint in $L^2(\euc^d)$, we have the solution $\psi(q,t)=e^{-i\hbar^{-1}tH}\underline{\psi}(q)$ by Stone's theorem \footnote{see, Stone's theorem in Reed-Simon \cite{RS72}}.
On the other hand,  Lie-Trotter-Kato's product formula\footnote{see, `` Trotter product formula'' in \cite{RS72}} asserts that even though $[H_0,V]\neq0$, we have
$$
e^{-i\hbar^{-1}tH}=e^{-i\hbar^{-1}t(H_0+V)}=\lim_{n\to\infty}(e^{-i\hbar^{-1}(t/n)V}e^{-i\hbar^{-1}(t/n)H_0})^n.
$$
Since
$$
e^{-i\hbar^{-1}tH_0}u(q)=(2\pi i\hbar t/m)^{-d/2}\int_{\euc^d}dq'\, e^{im(q-q')^2/(2\hbar t)}u(q'),
$$
we have
$$
e^{-i\hbar^{-1}(t/n)V} e^{-i\hbar^{-1}(t/n)H_0}u(q)=e^{-i\hbar^{-1}(t/n)V(q)}\int_{\euc^d}dq'\, e^{im(q-q')^2/(2\hbar t/n)}u(q'),
$$
we get, 
$$
(2\pi i\hbar t/m)^{-d/2}\int_{\euc^d}dq'\, e^{im(q-q')^2/(2\hbar t)}u(q')
=(2\pi i\hbar t/m)^{-dn/2}\int_{\euc^{dn}}dq_0{\cdots}dq_{n-1}e^{i\hbar S^*_t(q_n,\cdots,q_0)}u(q_0)
$$
with $q'=q_0$ and $q_n=q$, where
$$
S^*_t(q_n,\cdots,q_0)=
\sum_{j=1}^n\bigg[\frac{m}{2}\frac{(q_{j}-q_{j-1})^2}{(t/n)^2}-V(q_{j})\bigg]\frac{t}{n}.
$$
Taking $\gamma^*(\tau)$ on $[0,t]$ as a zigzag path from $q_0$ at time $0$, passing through $\gamma(\tau_j)=q_j$ $j=0,{\cdots},n$ where $\tau_j=j\frac{t}{n}$ and
$q_0,{\cdots}, q_n$ are given points in $\euc^d$. Feynman regards this $S^*_t(q_n,\cdots,q_0)$ as a Riemann approximation for the classical action $S_t(\gamma)$
 along the path $\gamma(\tau)$:
$$
S_t(\gamma)=\int_0^td\tau\,\bigg[\frac{m}{2}\bigg(\frac{d\gamma}{d\tau}\bigg)^2- V(\gamma(\tau))\bigg].
$$
Assuming $\gamma^*(\tau)$ is an approximation for any classical path belonging to the path space
$$
\Gamma_{(t,\unbq,\barq)}=\{\gamma\in AC([0,t]:\euc^d); \gamma(0)=\unbq, \gamma(t)=\barq\}
$$
and using assumed ``Lebesgue-like measure"$d_F\gamma$ on $\Gamma_{(t,\unbq,\barq)}$, we express its solution as
$$
\psi(\barq,t)=\int_{\gamma(t)=\barq}d_F\gamma\,e^{i\hbar^{-1} S_t(\gamma)}\underline{\psi}(\gamma(0)),
$$
called Feynman's path integral expression for the solution of Schr\"odinger equation by {Feynman's time-slicing method}.

If we permit this integral representation of the solution with operations under integral sign admitted, when making $\hbar\to 0$, we have the main contribution stems from the stationary point $\gamma_c(\cdot)$, that is, $\displaystyle{\frac{\delta S_t(\gamma)}{\delta \gamma}\bigg|_{\gamma=\gamma_c}}=0$.
This expression with above interpretation is persuasive to claim that classical mechanical equation appeared from quantum one when making $\hbar\to 0$.

\subsection{Quantization \`a la Fujiwara}
Using Feynman's idea rather conversely, Fujiwara \cite{fuj79,fuj80} 
constructed a fundamental solution of Schr\"odinger equation.
From a given Lagrangian $I(q,\dot{q})$, he found a classical orbit $\gamma_c(\cdot)\in \Gamma_{(t,\unbq,\barq)}$ of that Lagrangian mechanics:
$$
\ddot q(s)+I(q(s),\dot{q}(s))=0 \with q(0)=\unbq, \; \dot q(0)=\unbp^*
$$
such that $\barq=q(t,\unbq,\unbp^*)$, that is, $\gamma_c(0)=\unbq, \gamma_c(t)=\barq.$
Then, the action integral corresponding to $\gamma_c$ is given
$$
S_t(\gamma_c)=S_t(\barq,\unbq)=\int_0^t ds\,L(\gamma_c(s),\dot{\gamma}_c(s)),
$$
moreover defining van Vleck determinant as
$$
D(t,\barq,\unbq)=\det\bigg(\frac{\partial^2}{\partial \unbq \partial \barq}S_t(\barq,\unbq)\bigg),
$$
he finally defined a short time propagator as a FIOp(=Fourier Integral Operator)
$$
T_t\underline{\psi}(\barq)=\int_{\euc^d}d\unbq\,A(t,\barq,\unbq)e^{i\hbar^{-1}S_t(\barq,\unbq)}\underline{\psi}(\unbq)
$$
where
$$
A(t,\barq,\unbq)=\sqrt{D(t,\barq,\unbq)}.
$$

Above Fujiwara's process is justified when $\sup_q|\partial_q^\alpha V(q)|\leq C, (|\alpha|\geq2)$. That is, under this condition, not only there exists a unique classical orbit, and above quantities are well-defined, but also $T_t$ defines a bounded linear operator on $L^2(\euc^d)$ with the property
$$
\Vert T_{t+s}u-T_sT_tu\Vert\leq C(t^2+s^2)\Vert u\Vert.
$$
Moreover, following operator $E_t$ is defined by
$$
\lim_{n\to\infty} \Vert (T_{t/n})^n u-E_tu\Vert=0
$$
which gives a parametrix of the given initial  value problem of Schr\"odinger equation \cite{fuj79}.
The kernel representation is the fundamental solution such that
$$
(E_tu)(\barq)=\int_{\euc^d}\, d\unbq K(t,\barq,\unbq)u(\unbq).
$$
Finally, not only how to characterize the long-time behavior\footnote{For example, how Maslov index appeared in PIM?} of $K(t,\barq,\unbq)$ \cite{fuj80}, but also how to use the $\#$ products of FIOps by Kumano-go group \cite{KK81,kum76,KTT78}
are the problems reconsidered.

\begin{remark}
(1)[Feynman-Kac formula, see Simon \cite{sim79}]:
Stimulated by Feynman's idea, M. Kac represents the solution of the heat type equation
$$
\pdt u(q,t)=\sigma\Delta u(q,t)-V(q)u(q,t),
$$
using Wiener measure $dW(\gamma)$ by
$$
u(q,t)=\int dW(\gamma)\,e^{-\int_0^t V(\gamma(s)+q)ds}\unbu(\gamma(0)+\unbq). 
$$
(2)[Problem of $R/12$]: For a given Riemann manifold $(M, g_{jk})$, applying Fujiwara's idea to the heat equation, that is, some type of quantization, Inoue and Maeda \cite{IM85} got the notorious term $R/12$ where $R$ is the scalar curvature of the Riemann metric $g_{jk}(q)dq^jdq^k$. Curiously, this term was also derived as the most probable path calculation in probability theory, Fujita and Kotani \cite{FK82}, Takahashi and Watanabe \cite{TW80}.
\par
Rather recently the term $R/12$ appeared in Fukushima \cite{fuk21}, though he tries to construct a fundamental solution of given equation $\displaystyle{i\pdt-\frac{1}{2}\Delta_g+\frac{R}{12}}$. That means the origin of the term $R/12$ isn't explained there from the quantization process of $g_{jk}(q)dq^jdq^k$\footnote{an example of the so-called ordering problem} like physicists arguments, for example, DeWitt \cite{DW84} or Bastianelli, Corradini and Vassura \cite{BCV17}. 
\end{remark}

\subsection{Hamilton formulation of PIM}
For a given Lagrangian $I(q,\dot{q})$, above process is denoted formally as
$$
K=\int d_F\gamma\, e^{i\hbar^{-1} S(\gamma)}\quad\mbox{or}\quad K(t,\barq,\unbq)=\int_{\Gamma_{(t,\barq,\unbq)}} d_F\gamma\, e^{i\hbar^{-1} S(\gamma)}
$$
where
$$
S(\gamma)=\int_0^t ds \,L(\gamma(s),\dot{\gamma}(s)),{\;\;}\mbox{for each}\; \gamma\in\Gamma_{(t,\barq,\unbq)}.
$$
What does it become when Hamiltonian $H(q,p)$ is given? Here, as is well-known, $H(q,p)$ is related to $I(q,\dot{q})$ by Legendre transformation.
How to give meaning to the following formal expression?:
$$
K=\int d_Fp\, d_Fq\, e^{i\hbar^{-1}  \int ds(\dot{q}p-H(q,p))}.
$$
Under same assumption as above, there exists a classical orbit $(q(t,\unbq,\unbp),p(t,\unbq,\unbp))$ to the Hamilton equation for the given $H(q,p)$ 
with initial data $(\unbq,\unbp)$. For any fixed $\unbp$, for the map
$\unbq\to\barq=q(t,\unbq,\unbp)$, there exists an inverse map $\unbq=x(t,\barq,\unbp)$ for sufficiently small $|t|$:
$$
\barq=q(t,x(t,\barq,\unbp),\unbp) \;\et \; \unbq=x(t,q(t,\unbq,\unbp),\unbp)
$$
Using these, we put
$$
S(t,\barq,\unbp)=(\unbq\unbp-S_0(t,\unbq,\unbp))\bigg|_{\unbq=x(t,\barq,\unbp)}
$$
and define
$$
D(t,\barq,\unbp)=\det\bigg(\partial_{\barq}\partial_{\unbp}S(t,\barq,\unbp)\bigg).
$$
Finally we define
$$
T_tu(\barq)=(2\pi\hbar)^{-d/2}\int_{\euc^d} d\unbp\,D(t,\barq,\unbp)^{1/2}e^{i\hbar^{-1}S(t,\barq,\unbp)}\hat{u}(\unbp).
$$
This gives a parametrix for Schr\"odinger equation for short time. More precisely, see, Inoue \cite{ino99}.

\begin{remark} 
(1) In order to consider a classical orbit connecting 2 points $\barq$ and $\unbq$ in $t$, we need 2nd order time derivatives in corresponding classical mechanical equation.
Then, how one treats Weyl or Dirac equation?  Hint is given in the next section.
\newline
(2)[Hamiltonian formulation on manifolds?]
Above mentioned works containing physicist papers, are formulated in Lagrangian case. Since it needs works to formulate Fourier transformations on manifolds, physicist papers such as Field \cite{fie06, fie12}, claim something interesting without any mathematical estimates.
Seemingly to use directly Hamilton formulation in quantization as in \cite{ino99}, it is indispensable to define Fourier transformation even on general manifold, see Widom \cite{wid80}.
\end{remark}

\section{Another interpretation of method of characteristics} 
To start with this section, we check the difference between Lagrangian or Hamiltonian methods, by taking the simplest example.
This is important to treat first order system of PDE such as Dirac or Weyl equation by applying Feynman's time slicing method. 

We recall the method of characteristics. On region $\Omega\subset{\euc}^{d+1}$, we consider the following initial value problem: Let solve
\begin{equation}
\left\{
\begin{aligned}
&\pdt u(t,q)+\sum_{j=1}^d a_j(t,q){\pdqj}u(t,q)=b(t,q)u(t,q)+f(t,q),\\
&u({\underline{t}},q)={\underline{u}}(q)
\end{aligned}\right.
\label{mc7}
\end{equation}
The charactrisitic equation of \eqref{mc7} is given by
$$
\left\{
\begin{aligned}
&\frac d{dt} q_j(t)=a_j(t,q(t)),\\
& q_j({\underline{t}})={\underline{q}}_j
\for j=1,\cdots,d,
\end{aligned}
\right.
$$
with solution denoted by
$$
q(t)=q(t,{\underline{t}};{\underline{q}})=(q_1(t),\cdots,q_d(t))\in\euc^d, {\underline{q}}=({\underline{q}}_1,{\cdots},{\underline{q}}_d).
$$

Using this, we have
\begin{thm}[method of characteristics] \label{moc}
In \eqref{mc7}, assume coefficients $a_j\in C^1(\Omega:\euc)$, $b,f\in C(\Omega:\euc)$.
Taking any point $({\underline{t}},{\underline{q}})\in\Omega$,
assume ${\underline{u}}$ is $C^1$ in a neighbourhood of ${\underline{q}}$.
Then, there exists a unique solution $u(t,q)$ near $({\underline{t}},{\underline{q}})$. 
Moreover, putting $B(t,{\underline{q}})=b(t,q(t,{\underline{t}};{\underline{q}}))$, 
$F(t,{\underline{q}})=f(t,q(t,{\underline{t}};{\underline{q}}))$ and defining
\begin{equation}
U(t,{\underline{q}})
=e^{\int_{{\underline{t}}}^t d\tau\, B(\tau,{\underline{q}})}
\left\{\int_{{\underline{t}}}^t ds\,
e^{-\int_{{\underline{t}}}^s d\tau\, B(\tau,{\underline{q}})}
F(s,{\underline{q}})+
{\underline{u}}({\underline{q}})\right\}
\label{mc8}
\end{equation}
we have a solution \eqref{mc7} given by
\begin{equation}
u(t,\bar{q})=U(t,x(t,{\underline{t}};{\overline{q}})).
\label{mc8-1}
\end{equation}
Here, ${\underline{q}}=x(t,{\underline{t}};{\overline{q}})$ is the inverse function of
${\bar{q}}=q(t,{\underline{t}};{\underline{q}})$.
\end{thm}
\begin{remark}
\eqref{mc8} satisfies
$$
\dt U(t,{\underline{q}})
=B(t,{\underline{q}})U(t,{\underline{q}})+F(t,{\underline{q}}).
$$
\end{remark}

\paragraph{\bf An example:} With this theorem in mind, we consider the following simplest case:
\begin{equation}
\left\{
\begin{aligned}
&i\hbar\pdt u(t,q)=a\frachi{\pdq} u(t,q)+bqu(t,q),\\
&u(0,q)={\underline{u}}(q).
\end{aligned}\right.
\label{SE1}
\end{equation}
The symbol of the righthand side of above equation is derived by
$$
H(q,p)=e^{-i\hbar^{-1}qp}\bigg(a\frachi{\pdq}+bq\bigg)e^{i\hbar^{-1}qp}=ap+bq.
$$
From this, we have the characteristic
$$
\dot q(t)=\frac{\partial H(q,p)}{\partial p}=a
\with
q(0)=\underline{q},
$$
whose solution is given
$$
q(s)={\underline{q}}+as 
\with {\underline{q}}=x(t,{\overline{q}})={\overline{q}}-at.
$$
Since above representation \eqref{mc8} gives
$$
U(t,{\underline{q}})=
{\underline{u}}({\underline{q}})
e^{-i\hbar^{-1}(b{\underline{q}}t+2^{-1}ab{t^2})},
$$
from \eqref{mc8-1}, we get
$$
u(t,{\overline{q}})={\underline{u}}({\overline{q}}-at)
e^{-i\hbar^{-1}(b{\overline{q}}t-2^{-1}ab{t^2})}.
$$
By this procedure, the information of $p(t)$ isn't used!

Now, we give another interpretation of this representation by Hamilton path integral method via Hamilton-Jacobi equation.

For $(t,{\underline{q}},{\underline{p}})$, putting
$$
S_0(t,{\underline{q}},{\underline{p}})
=\int_0^t ds [\dot q(s)p(s)-H(q(s),p(s))]=-b{\underline{q}}t-2^{-1}ab{t^2},
$$
we define the action integral for Hamilton function $H(q,p)$
$$
S(t,{\overline{q}},{\underline{p}})
={\underline{q}}\,{\underline{p}}+S_0(t,{\underline{q}},{\underline{p}})
\big|_{{\underline{q}}=x(t,{\overline{q}})}={\overline{q}}{\underline{p}}-a{\underline{p}}t-b{\overline{q}}t
+2^{-1}ab{t^2}.
$$
This $S=S(t,{\overline{q}},{\underline{p}})$ satisfies the following Hamilton-Jacobi equation:
$$
{\pdt}S+H({\overline{q}},\partial_{\overline{q}}S)=0 \with S(0,{\overline{q}},{\underline{p}})={\overline{q}}{\underline{p}}.
$$
In this case, van Vleck determinant is a scalar
$$
\frac{\partial^2S(t,{\overline{q}},{\underline{p}})}{\partial{\overline{q}}\partial{\underline{p}}}=1
$$
and it satisfies the following continuity equation:
$$
{\pdt}D+\frac{1}{2}{\partial_{\overline{q}}}(DH_p)=0 \with
D(0,{\overline{q}},{\underline{p}})=1\where H_p=\frac{\partial H}{\partial p}({\overline{q}},\partial_{\overline{q}}S).
$$
Using these classical quantities $S$ and $D$, modifying Feynman's method slightly\footnote{Though Feynman's procedure based on Lagrangian, here we changed to Hamiltonian formulation}, we define
$$
u(t,{\overline{q}})=(2\pi\hbar)^{-1/2}\int d{\underline{p}}\,
D^{1/2}(t,{\overline{q}},{\underline{p}}){\cdot}e^{i\hbar^{-1}S(t,{\overline{q}},{\underline{p}})}\hat{\underline{u}}({\underline{p}}),
$$
Using rather formal expression $\displaystyle{\delta(\overline{q}-at-\underline{q})=(2\pi\hbar)^{-1}\int d{\underline{p}}\, e^{i\hbar^{-1}(\overline{q}-at-\underline{q}){\underline{p}}}}$, we have
$$
\begin{aligned}
u(t,{\overline{q}})&=(2\pi\hbar)^{-1/2}\int d{\underline{p}}\,e^{i\hbar^{-1}S(t,{\overline{q}},{\underline{p}})}\hat{\underline{u}}({\underline{p}})\\
&=(2\pi\hbar)^{-1}\iint d{\underline{p}}d{\underline{q}}\,
e^{i\hbar^{-1}(S(t,{\overline{q}},{\underline{p}})-\underline{q}{\underline{p}})}{\underline{u}}({\underline{q}})
=\int d{\underline{q}}\,\delta({\overline{q}}-at-{\underline{q}}){\underline{u}}(\underline{q})
e^{i\hbar^{-1}(-b{\overline{q}}t+2^{-1}ab{t^2})}\\
&={\underline{u}}({\overline{q}}-at)
e^{i\hbar^{-1}(-b{\overline{q}}t+2^{-1}ab{t^2})}.
\end{aligned}
$$

\begin{problem} 
How we interpret the solution of \eqref{mc7} by this idea? More generally, how to apply this interpretation to systems of PDE, for example, Weyl or Dirac equations\footnote{See, Inoue \cite{ino98-1,ino98-2,ino99-3}.}?
\end{problem}

\section{Necessity of the new dimension} 
\subsection{Reasons of necessity}
There are several reasons at least for me except the claim of Manin \cite{man85} or Gromov \cite{grom99}: 
(i) As is written in p.355 of \cite{FH65}, 
\underline{`spin' has been the object outside Feynman's procedures} at that time (below, underlined by atlom):
\begin{quote} 
$\cdots $ \underline{path integrals suffer grievously from a serious defect}.
They do not permit a discussion of spin operators or other such operators 
in a simple and lucid way.
They find their greatest use in systems for which coordinates and their
conjugate momenta are adequate.
Nevertheless, spin is a simple and vital part of real quantum-mechanical
systems.
It is a serious limitation that the half-integral spin of the electron
does not find a simple and ready representation.
It can be handled if the amplitudes and quantities are considered
as quarternions instead of ordinary complex numbers,
but the lack of commutativity of such numbers is a serious complication.
\end{quote}
(ii) Berezin, who invented the second quantization,  proposed with Marinov \cite{BM77} 
to ``treat boson and fermion on equal footing''. That is, instead of $\euc$ or $\mathbb{C}$, he claims the necessity to construct a scalar-like, non-commutative ground field, where lives electron and photon equally.\\
(iii) Witten \cite{witt82-1}
explained the meaning of supersymmetric idea in physics to mathematician by new derivation of Morse inequalities using exterior differential operations. But be careful, in many case, physicists usage of supersymmetry only means to use fermion or odd variables, not exactly treating some ``symmetry'' in ${\fR}^{m|n}$, for example Efetov's works \cite{efe83, efe05}, 
Brezin \cite{bre85} and Fyodorov \cite{fyo95}. 

\subsection{Regarding matrices as differential operators}
A claim ``any Clifford algebra has a representation on Grassmann algebra"\footnote{Though I don't appreciate fully Chevalley' theorem``Any Clifford algebra has a representation on Grassmann algebra", but following  consideration may be sufficient of my intuitive arguments.} is straight forwardly explained for $2\times 2$ matrices case.
Though analogous argument works for a set of $2^\ell\times 2^\ell$ matrices because it has Clifford relation, but for a set of general $N\times N$ matrices, we need to find suitable algebra on which we need foundation of analysis. I imagine  not only the work of Martin \cite{mar59-1, mar59-2} but also the one by Campoamor-Stursberg  and Rausch de Traubenberg \cite{CSRT09}
give some hint on this.
Especially $N=3$ will be interesting when we consider vorticity equation for Euler or Navier-Stokes equation which is represented by $3\times 3$ matrices. (Recall the construction of the classical solution of Euler equation by Kato \cite{kat67} 
using method of characteristics since vorticity equation is reduced to scalar one when the space dimension is $2$.)

Recall Pauli matrices $\{\pmb{\sigma}_k\}_{k=1}^3$:
$$
\pmb{\sigma}_1=\begin{pmatrix}
0&1\\
1&0
\end{pmatrix},\quad
\pmb{\sigma}_2=-i\begin{pmatrix}
0&1\\
-1&0
\end{pmatrix},\quad
\pmb{\sigma}_3=\begin{pmatrix}
1&0\\
0&-1
\end{pmatrix}.
$$
These matrices satisfy not only Clifford relation
\begin{equation}
\pmb{\sigma}_j\pmb{\sigma}_k+\pmb{\sigma}_k\pmb{\sigma}_j=2\delta_{jk}, \where j,k=1,2,3,
\label{CR}
\end{equation}
but also the following for any $(j,k,\ell)$, an even permutation of $(1,2,3)$:
\begin{equation}
\pmb{\sigma}_j\pmb{\sigma}_k=i\pmb{\sigma}_\ell.
\label{CR2}
\end{equation}

Decompose any $2\times 2$ matrix $\mathbb\hbar^{-1}$ by $\{\pmb{\sigma}_j\}$,
\begin{equation}
\begin{aligned}
{\mathbb\hbar^{-1}}=\begin{pmatrix}
a&c\\
d&b
\end{pmatrix}
&=\frac{a+b}{2}\begin{pmatrix}
1&0\\
0&1
\end{pmatrix}
+\frac{a-b}{2}\begin{pmatrix}
1&0\\
0&-1
\end{pmatrix}\\
&\qquad\qquad
+\frac{c+d}{2}\begin{pmatrix}
0&1\\
1&0
\end{pmatrix}
+\frac{c-d}{2}\begin{pmatrix}
0&1\\
-1&0
\end{pmatrix}\\
&=\frac{a+b}{2}\mathbb{I}_2+\frac{a-b}{2}\pmb{\sigma}_3+\frac{c+d}{2}\pmb{\sigma}_1+i\frac{c-d}{2}\pmb{\sigma}_2.
\end{aligned}
\label{MRbyP}
\end{equation}
That is,  a set of all $2\times 2$ matrices has the Clifford algebra structure with Pauli matrices $\{\pmb{\sigma}_k\}$ as a basis.

We identify a vector as a function of Grassmann variables. Then, a matrix is regarded as a differential operator acting on a set of functions composed with Grassmann variables, which forms something-like field similar to real or complex number field but non-commutative.

Rather abruptly we prepare\footnote{usage of the word ``preparing'' odd variables $\theta_1$, $\theta_2$ is not popularly used in PDE group} two odd variables  $\theta_1, \theta_2$ having the following relations:
\begin{equation}
\theta_1{\cdot}\theta_2+\theta_2{\cdot}\theta_1=0, \; \theta_j{\cdot}\theta_j=0, (j=1,2)
\label{GR}
\end{equation}
That is, they have product ${\cdot}$ having Grassmann relations and differentiation such as
$$
\frac{\partial}{\partial \theta_1}1=0, \quad \frac{\partial}{\partial \theta_1}\theta_1=1, \quad etc.
$$
Here, when we differentiate, bringing that variable in front, i.e.
$$
\frac{\partial}{\partial \theta_2}\theta_1{\cdot}\theta_2=\frac{\partial}{\partial \theta_2}(-\theta_2{\cdot}\theta_1)
=\frac{\partial}{\partial \theta_2}(\theta_2{\cdot}(-\theta_1))=-\theta_1.
$$

More concretely, for example, taking  two variables $z_1$ and $z_2$ in $\euc^2$, construct differential forms $dz_1$, $dz_2$ with exterior  product $\wedge$ and interior product $\lfloor$ such that
$$
dz_1\wedge dz_2=-dz_2\wedge dz_1, \quad dz_j\wedge dz_j=0,\quad \frac{\partial}{\partial{z_j}}\lfloor dz_k=\delta_{jk}.
$$
Here, identifying 
$dz_j$ as $\theta_j$ for $j=1,2$, $\wedge$ as ${\cdot}$ and $\frac{\partial}{\partial{z_j}}\lfloor$ as $\frac{\partial}{\partial \theta_j}$, and abbreviating ${\cdot}$, we continue to explain.

Using odd variables $\theta_1$, $\theta_2$, we give the identifying maps $\#$, $\flat$ with Grassmann algebras as follows:
\begin{equation}
\begin{gathered}
\Gamma_0=\{u(\theta)=u_0+u_1\theta_1\theta_2 \;|\; u_0, u_1\in{\mathbb C}\}
\,\begin{matrix}{\overset{\flat}{\rightarrow}}\\[-8pt]
{\underset{\sharp}{\leftarrow}}\end{matrix}\;
{\mathbb C}^2=\bigg\{{\bf u}=\binom{u_0}{u_1}\;\big|\; u_0, u_1\in{\mathbb C} \bigg\}\\
\with
(\#{\bf u})(\theta)=
\bigg(\#
\begin{pmatrix}
u_0\\
u_1
\end{pmatrix}\bigg)(\theta)
=u_0+u_1\theta_1\theta_2,\\
\flat(u_0+u_1\theta_1\theta_2)=\flat (u(\theta))=\begin{pmatrix}
u_0\\
u_1
\end{pmatrix}. 
\end{gathered}
\label{sharp-flat}
\end{equation}
Here,  we have $u(0)=u(\theta)|_{\theta=0}=u_0$, $\partial_{\theta_2}\partial_{\theta_1}u(\theta)\big|_{\theta=0}
=\partial_{\theta_2}(\partial_{\theta_1}u(\theta))|_{\theta=0}=u_1$, which relates a vector $\bf u$ and a function $u(\theta)$.

Now, we define differential operators w.r.t. odd variables:
\begin{equation}
\begin{aligned}
&\sigma_1(\theta,\partial_\theta)=
\theta_1\theta_2
-\frac{\partial^2}{\partial\theta_1\partial\theta_2},\\
&\sigma_2(\theta,\partial_\theta)=i\bigg(\theta_1\theta_2
+\frac{\partial^2}{\partial\theta_1\partial\theta_2}\bigg),\\
&\sigma_3(\theta,\partial_\theta)=
1-\theta_1\frac{\partial}{\partial\theta_1}
-\theta_2\frac{\partial}{\partial\theta_2}.
\end{aligned}\label{1.7}
\end{equation}
Thes operators act on $u(\theta)=u_0+u_1\theta_1\theta_2\in\Gamma_{0}$ as follows:
$$
\begin{aligned}
&\sigma_1(\theta,\partial_\theta)(u_0+u_1\theta_1\theta_2)=u_0\theta_1\theta_2+ u_1,\\
&\sigma_2(\theta,\partial_\theta)(u_0+u_1\theta_1\theta_2)=i(u_0\theta_1\theta_2- u_1),\\
&\sigma_3(\theta,\partial_\theta)(u_0+u_1\theta_1\theta_2)=u_0-u_1\theta_1\theta_2.
\end{aligned}
$$
Here,
$$
\begin{aligned}
\frac{\partial^2}{\partial{\theta_1}\partial{\theta_2}}\theta_1\theta_2
&=\partial_{\theta_1}\partial_{\theta_2}\theta_1\theta_2\\
&=\partial_{\theta_1}(\partial_{\theta_2}\theta_1\theta_2)
=\partial_{\theta_1}(\partial_{\theta_2}(-\theta_2\theta_1))=\partial_{\theta_1}(-\theta_1)=-1.
\end{aligned}
$$

They act on $\Gamma_0$ as,
$$
\sigma_1(\theta,\partial_\theta)\sigma_2(\theta,\partial_\theta)
=i\sigma_3(\theta,\partial_\theta),\quad
\sigma_2(\theta,\partial_\theta)\sigma_3(\theta,\partial_\theta)
=i\sigma_1(\theta,\partial_\theta),\quad
\sigma_3(\theta,\partial_\theta)\sigma_1(\theta,\partial_\theta)
=i\sigma_2(\theta,\partial_\theta).
$$
\par
\begin{remark}
(1) The action $\sigma_1(\theta,\partial_\theta)$ on  $\Gamma_0$ has a matrix representation
$$
\flat \sigma_1(\theta,\partial_\theta) \sharp \begin{pmatrix}
u_0\\
u_1
\end{pmatrix}=\begin{pmatrix}
u_1\\
u_0
\end{pmatrix},\quad
\flat \sigma_1(\theta,\partial_\theta) \sharp=\pmb{\sigma}_1, \quad etc.
$$
(2) Moreover, putting
$$
\Gamma_1=\{v(\theta)=v_1\theta_1+v_2\theta_2\;|\; v_1, v_2\in{\mathbb C}\}
\,\begin{matrix}{\overset{\flat}{\rightarrow}}\\[-8pt]
{\underset{\sharp}{\leftarrow}}\end{matrix}\;
{\mathbb C}^2=\bigg\{{\bf v}=\binom{v_1}{v_2}\;\big|\; v_1, v_2\in{\mathbb C} \bigg\}
$$
we have
$$
\sigma_j(\theta,\partial_\theta)v(\theta)=0 \forany  v\in \Gamma_1\et j=1,2,3.
$$
\end{remark}

We define Fourier transformation\footnote{The integration here is very algebraic, called Berezin integral, which has properties as integral. See, more precisely, Vladimirov and Volovich \cite{VV83, VV84}.} for odd variables\footnote{We need to remark that the definition of supernumber field we use here, is different from above related works because they are formed by countably infinite number of Grassmann generators, i.e. on Fr\'echet-Grassmann algebra! See, Inoue \cite{ino14-1, ino14-2}, which are largely revised version of \cite{ino92}}.
Taking number ${\spin}$ in ${\euc}^{\times}={\euc}-\{0\}$ or $i{\euc}^{\times}$ called it as spin constant, we define
$$
\hat{u}(\pi)={\spin}\int_{{\fR}^{0|2}}d\theta\,
e^{-i{\spin}^{-1}\langle\theta|\pi\rangle}u(\theta),\quad
{u}(\theta)={\spin}\int_{{\fR}^{0|2}}d\pi\,
e^{i{\spin}^{-1}\langle\theta|\pi\rangle}\hat{u}(\pi).
$$
In the above, the integral interval ${\fR}^{0|2}$ may be considered only as a symbol, at least for the time being.
Here, $\langle\theta|\pi\rangle=\theta_1\pi_1+\theta_2\pi_2$ and
$$
\langle\theta|\pi\rangle^2=2\theta_1\pi_1\theta_2\pi_2=-2\theta_1\theta_2\pi_1\pi_2,\quad 
\langle\theta|\pi\rangle^j=\overbrace{\langle\theta|\pi\rangle\langle\theta|\pi\rangle{\cdots}\langle\theta|\pi\rangle}^{\text{$j$ times}}=0\;\mbox{if}\; j\ge 3
$$
we have
$$
e^{a\langle\theta|\pi\rangle}=\sum_{{\ell}=1}^{\infty}\frac{a^{\ell}\langle\theta|\pi\rangle^{\ell}}{{\ell}!}
=1+a\langle\theta|\pi\rangle-a^2\theta_1\theta_2\pi_1\pi_2.
$$
\begin{remark}
``Integration'' should be considered preferable such that (i) all polynomials are integrable, (ii) linear with integrand and (iii) translation invariant, Berezin integration satisfies these properties.
\end{remark}
By Fourier transformation, differentiation is regarded as multiplication in dual space\footnote{Does this mean that the four basic arithmetic operations is the only core to recognize objects mathematically?}, and theory of pseudo-differential operators treats PDE with variable coefficients controlling error estimates.

Defining Weyl symbols of differential operators $\{\sigma_j(\theta,\partial_{\theta})\}_{j=1}^3$ as
\begin{equation}
\left\{\begin{aligned}
&\sigma_1(\theta,\pi)=\theta_1\theta_2+{\spin}^{-2}\pi_1\pi_2,\\
&\sigma_2(\theta,\pi)=i(\theta_1\theta_2-{\spin}^{-2}\pi_1\pi_2),\\
&\sigma_3(\theta,\pi)=-i{\spin}^{-1}\langle\theta|\pi\rangle
=-i{\spin}^{-1}(\theta_1\pi_1+\theta_2\pi_2),
\end{aligned}\right.
\label{WS}
\end{equation}
we have, for example,
$$
\begin{aligned}
(\hat{\sigma}_3^{\mathrm{w}}(\theta,\partial_{\theta})u)(\theta)&=-i{\spin}\iint_{{\fR}^{0|2}\times {\fR}^{0|2}}d\pi d\theta'\, e^{i{\spin}^{-1}\langle\theta-\theta'|\pi\rangle} \sigma_3(\frac{\theta+\theta'}{2},\pi)
(u_0+u_1{\theta'}_1{\theta'}_2)\\
&=u_0-u_1\theta_1\theta_2=\sigma_3(\theta,\partial_{\theta})(u_0+u_1\theta_1\theta_2).\;\; etc.
\end{aligned}
$$
Here, we used
$$
\int_{{\fR}^{0|2}}d\pi\, e^{i{\spin}^{-1}\langle\theta-\theta'|\pi\rangle} \sigma_3(\frac{\theta+\theta'}{2},\pi)
=-i\kbar^{-1}(\theta_1\theta_2-\theta_1'\theta_2').
$$
\begin{remark} 
(i) It takes many time to perceive the meaning of $1$ in ${\sigma}_3(\theta,\partial_{\theta})$ appeared in \eqref{1.7}, that is, $1$ stems from
Weyl quantization!\\
(ii) Artificially introduced spin constant $\spin$ gives nice result only when $\spin\hbar^{-1}=1$, that is, the action integral obtained by Jacobi method satisfies Hamilton-Jacobi equation in case $\spin\hbar^{-1}=1$, therefore we assume $\spin=\hbar$!
 \end{remark}

Therefore, from \eqref{MRbyP}, we have
\begin{equation}
({\sharp}{\mathbb\hbar^{-1}}{\bf u})(\theta)=\bigg[\frac{a+b}{2}+\frac{a-b}{2}{\sigma}_3(\theta,\partial_{\theta})+\frac{c+d}{2}{\sigma}_1(\theta,\partial_{\theta})+i\frac{c-d}{2}{\sigma}_2(\theta,\partial_{\theta})\bigg]u(\theta),
\end{equation}
whose Weyl symbol is
\begin{equation}
\sigma^{\rm w}({\sharp}{\mathbb\hbar^{-1}}{\flat})=\frac{a+b}{2}+\frac{a-b}{2}{\sigma}^{\rm w}_3(\theta,\pi)+\frac{c+d}{2}{\sigma}_1(\theta,\pi)+i\frac{c-d}{2}{\sigma}_2(\theta,\pi).
\end{equation}

\begin{remark}
In the theory of pseudo-differential operators, symbol calculus is the main ingredient where algebra is essentially $\euc$.
Whether analogous procedure works on superspace ${\fR}^{m|n}$?
More explicitly, though the symbol of $\sigma_1(\theta,\partial_\theta)$ is given as $\theta_1\theta_2+\kbar^{-2}\pi_1\pi_2$, but whether the inverse of $\sigma_1(\theta,\partial_\theta)$ is calculated directly from $\theta_1\theta_2+\kbar^{-2}\pi_1\pi_2$?
\end{remark}

\subsection{Chi's example}
Consider the following initial value problem
\begin{equation}
\partial_t^2u-L(t,q,\partial_q)u=f \with u(0,q)={\unbu}_0(q),\quad u_t(0,q)={\unbu}_1(q).
\label{chi0}
\end{equation}
The following result is a part of Chi \cite{chi58} in 1958: 
\begin{thm}[An example of weakly hyperbolic equation]
Let 
$$
L(t,q,\partial_q)=t^2\partial_q^2+b\partial_q.
$$
Solving the above initial value problem \eqref{chi0} with data $f=0$ and ${\unbu}_1=0$
when $b=4k+1$, $k\in {\mathbb{Z}}_+$, we have the solution given by
$$
u(t,q)=\sum_{\ell=0}^k\frac{2^{2\ell}\,k!}{(2\ell)!(k-\ell)!}t^{2\ell}{\unbu}_0^{(\ell)}\bigg(q+\frac{t^2}{2}\bigg).
$$
\end{thm}
\par
In this section, we derive this result by completely different method from Chi.

\subsubsection{Chi's proof}
Applying change of variables,
$$
\xi=x+\frac{t^2}{2},\quad \eta=x-\frac{t^2}{2}
$$
to $u(t,x)$, Chi reduces above problem to the Euler-Darboux equation for $\displaystyle{\tilde{u}(\xi,\eta)=u\bigg(\sqrt{\xi-\eta},\frac{\xi+\eta}{2}\bigg)}$ yielding,
\begin{equation}
\frac{\partial^2 \tilde{u}}{\partial \xi \partial \eta}-\frac{1-b}{4(\xi-\eta)}\frac{\partial \tilde{u}}{\partial\xi}+\frac{1+b}{4(\xi-\eta)}\frac{\partial \tilde{u}}{\partial\eta}=0
\label{chiED}
\end{equation}
with initial condition
$$
\tilde{u}(\xi,\xi)=\unbu_0(\xi),\quad \lim_{\xi-\eta\to0}(\xi-\eta)^{1/2}\bigg(\frac{\partial \tilde{u}}{\partial\xi}-\frac{\partial \tilde{u}}{\partial\eta}\bigg)=\unbu_1(\xi).
$$
\begin{remark}
For
$$
{E}(\alpha,\beta)=\partial_{\xi}\partial_{\eta}-\frac{\beta}{\xi-\eta}\partial_{\xi}+\frac{\alpha}{\xi-\eta}\partial_{\eta},
$$
with $\alpha=\frac{1+b}{4}, \beta=\frac{1-b}{4}$, it gives \eqref{chiED}.
He uses the fact such that if $0<\alpha, \beta<1$
then the solution of $E(\alpha,\beta)u=0$ is expressed as the convergent Euler-Darboux integral
$$
\begin{aligned}
u(\xi,\eta)&=\frac{\Gamma(\alpha+\beta)}{\Gamma(\alpha)\Gamma(\beta)}\int_0^1dt\,u_0(\xi+(\eta-\xi)t)t^{\beta-1}(1-t)^{\alpha-1}\\
&\qquad
+\frac{\Gamma(1-\alpha-\beta)}{2\Gamma(1-\alpha)\Gamma(1-\beta)}(\eta-\xi)^{t-\alpha-\beta}\int_0^1dt\,u_1(\xi+(\eta-\xi)t)t^{-\alpha}(1-t)^{-\beta}.
\end{aligned}
$$
Moreover, for other $\alpha,\beta$, this integral diverges as it stands, he gives these as Riemann-Liouville integral.
\par
Seemingly, Euler-Darboux equation has some relations to special functions or Painleve's functions,
therefore proceeding inversely, we might dream another perspective to these functions by using superanalysis?
\end{remark}

\subsubsection{Direct construction of a solution of Hamilton-Jacobi equation}
Without reducing to Euler-Darboux equation, we give a simple minded proof (at least as a story) of Chi's equation applying superanalysis (see also, \cite{ino24-1, ino24-2}). 

Putting $u_0=u$, $u_1=u_t$ and assuming for the moment $f=0$, we make \eqref{chi0} to a system :
$$
i\pdt\binom{u_0}{u_1}=i\begin{pmatrix}
0&1\\
L(t,\partial_q)&0
\end{pmatrix}
\binom{u_0}{u_1}.
$$
Here, we multiply $i$ to both sides, rather artificially.
Changing a vector representation to non-commutative scalar one by putting
$u(t,x,\theta)=u_0(t,x)+u_1(t,x)\theta_1\theta_2$, we have
$$
(L(t,\partial_x)\theta_1\theta_2-\partial_{\theta_1}\partial_{\theta_2})(u_0+u_1\theta_1\theta_2)
=u_1+Lu_0\theta_1\theta_2\sim \begin{pmatrix}
0&1\\
L(t,\partial_x)&0
\end{pmatrix}
\binom{u_0}{u_1}.
$$
Here, $q\in\euc$ is imbedded in $x\in{\fR}^{1|0}$ such that $X_{\rm B}=q$.
Defining
$$
{\mathcal{H}}(t,\partial_x,\theta,\partial_{\theta})
=iL(t,\partial_x)\theta_1\theta_2-i\partial_{\theta_1}\partial_{\theta_2}
$$
whose symbol is given by
$$
{\mathcal{H}}(t,\xi,\theta,\pi)=i(-t^2\xi^2+ib\xi)\theta_1\theta_2+i\pi_1\pi_2
$$
and corresponding Hamilton-Jacobi equation is given by
\begin{equation}
{\mathcal{S}}_t+{\mathcal{H}}(t,{\mathcal{S}}_x,\theta,{\mathcal{S}}_{\theta})=0
\label{HJ-chi}
\end{equation}
which is solved with the initial condition ${\mathcal{S}}(0,x,\xi,\theta,\pi)=\langle x|\xi\rangle +\langle\theta|\pi\rangle$.

In the supersmooth category, since we have
\begin{equation}
\begin{aligned}
{\mathcal{S}}(t,x,\xi,\theta,\pi)=&S(t,x,\xi)
+X(t,x,\xi)\theta_1\theta_2+Y(t,x,\xi)\theta_1\pi_1+\tilde{Y}(t,x,\xi)\theta_2\pi_2\\
&\qquad\qquad\quad+V(t,x,\xi)\theta_1\pi_2+\tilde{V}(t,x,\xi)\theta_2\pi_1\\
&\qquad\qquad\qquad+Z(t,x,\xi)\pi_1\pi_2+W(t,x,\xi)\theta_1\theta_2\pi_1\pi_2,
\end{aligned}
\label{QiS}
\end{equation}
where
$$
\begin{aligned}
S(t,x,\xi)&={\mathcal{S}}(t,x,\xi,0,0),\\
X(t,x,\xi)&=\partial_{\theta_2}\partial_{\theta_1}{\mathcal{S}}(t,x,\xi,\theta,\pi)\big|_{\theta=\pi=0},\\
Y(t,x,\xi)&=\partial_{\pi_1}\partial_{\theta_1}{\mathcal{S}}(t,x,\xi,\theta,\pi)\big|_{\theta=\pi=0},\quad
\tilde{Y}(t,x,\xi)=\partial_{\pi_2}\partial_{\theta_2}{\mathcal{S}}(t,x,\xi,\theta,\pi)\big|_{\theta=\pi=0},\\
V(t,x,\xi)&=\partial_{\pi_2}\partial_{\theta_1}{\mathcal{S}}(t,x,\xi,\theta,\pi)\big|_{\theta=\pi=0},\quad
\tilde{V}(t,x,\xi)=\partial_{\pi_1}\partial_{\theta_2}{\mathcal{S}}(t,x,\xi,\theta,\pi)\big|_{\theta=\pi=0},\\
Z(t,x,\xi)&=\partial_{\pi_2}\partial_{\pi_1}{\mathcal{S}}(t,x,\xi,\theta,\pi)\big|_{\theta=\pi=0},\quad
W(t,x,\xi)=\partial_{\pi_2}\partial_{\pi_1}\partial_{\theta_2}\partial_{\theta_1}{\mathcal{S}}(t,x,\xi,\theta,\pi)\big|_{\theta=\pi=0},
\end{aligned}
$$
we should seek solution ${\mathcal{S}}$ of \eqref{HJ-chi} with this form. 
Since
$$
{\mathcal{H}}(t,{\mathcal{S}}_x,\theta,{\mathcal{S}}_{\theta})
=i(-t^2{\mathcal{S}}_x^2+ib{\mathcal{S}}_x)\theta_1\theta_2+i{\mathcal{S}}_{\theta_1}{\mathcal{S}}_{\theta_2},
$$
putting $\theta=\pi=0$ in \eqref{HJ-chi}, we have readily
$$
S_t(t,x,\xi)=0 \with S(0,x,\xi)=\langle x|\xi\rangle.
$$
This gives $S(t,x,\xi)=\langle x|\xi\rangle$.

Differentiating \eqref{HJ-chi} w.r.t. $\theta_1$ and $\theta_2$ then restricting to $\theta=\pi=0$,
we get 
$$
X_t+i(-t^2\xi^2+i b\xi)+iX^2=0.
$$
Since this is the Riccati type ODE, using $\varphi(t)$, we find $X=-i\frac{\dot{\varphi}}{\varphi}$, that is,
\begin{equation}
\ddot{\varphi}+(t^2\xi^2-ib\xi)\varphi=0\with \dot{\varphi}(0)=0.
\label{chi-pow}
\end{equation}
Regarding $(x,\xi)$ as parameter, we may solve this equation by power series in $t$, but this procedure is postponed until explaining  our construction of ${\mathcal{S}}$ and quantization.

Putting $Y(t,x,\xi)=\partial_{\pi_1}\partial_{\theta_1}{\mathcal{S}}(t,x,\xi,\theta,\pi)\big|_{\theta=\pi=0}$, we have from \eqref{HJ-chi},
$$
Y_t+iXY=0\with Y(0)=1.
$$
From above structure of $X$, we have $Y\varphi=1$. Analogously, we have $\tilde{Y}(t,x,\xi)=\partial_{\pi_2}\partial_{\theta_2}{\mathcal{S}}(t,x,\xi,\theta,\pi)\big|_{\theta=\pi=0}$ which equals to $Y$.

Calculating $V$ and $\tilde{V}$ analogously, we get both equal to $0$.

As $Z(t,x,\xi)=\partial_{\pi_2}\partial_{\pi_1}{\mathcal{S}}(t,x,\xi,\theta,\pi)\big|_{\theta=\pi=0}$ satisfies
$$
Z_t+iY^2=0 \with Z(0)=0,
$$
we have
$$
Z(t,x,\xi)=-i\int_0^t ds\,Y^2(s)=-i\int_0^t ds\,\varphi(s)^{-2}.
$$

Analogously, we get $W=0$.

Therefore using $\varphi$, we have
\begin{equation}
{\mathcal{S}}=\langle x|\xi\rangle+X\theta_1\theta_2+Y\langle\theta|\pi\rangle+Z\pi_1\pi_2
\with
X=-i\frac{\dot{\varphi}}{\varphi},\quad Y=\frac{1}{\varphi}, \quad Z=-i\int_0^t ds\,\varphi(s)^{-2}.
\label{solHJ0}
\end{equation}

\subsubsection{Continuity equation}
Define van Vleck determinant as
$$
{\mathcal{D}}(t,x,\theta,\xi,\pi)
=\sdet
\begin{pmatrix}
\frac{\partial^2{\mathcal{S}}(t,x,\theta,\xi,\pi)}
{\partial x \partial\xi}
&\frac{\partial^2{\mathcal{S}}(t,x,\theta,\xi,\pi)}
{\partial x \partial\pi}\\
\frac{\partial^2{\mathcal{S}}(t,x,\theta,\xi,\pi)}
{\partial\theta \partial\xi}
&\frac{\partial^2{\mathcal{S}}(t,x,\theta,\xi,\pi)}
{\partial\theta \partial\pi}
\end{pmatrix},
$$
and abbreviated as ${\mathcal{D}}={\mathcal{D}}(t,x,\theta,\xi,\pi)$.
Using \eqref{solHJ0}, we have
$$
{\mathcal{D}}=\varphi^{2}.
$$

\subsubsection{Quantization}
Using $\mathcal{S}$ and $\mathcal\hbar^{-1}={\mathcal{D}}^{1/2}$, we define
\begin{equation}
\begin{aligned}
{\mathcal{T}}_t\unbu(\theta)&=
(2\pi)^{-1/2}\int d\xi\,d\pi\,{\mathcal\hbar^{-1}}e^{i{\mathcal{S}}}\hat{\unbu}(\pi)\\
&=(2\pi)^{-1/2}\int d\xi\,e^{i{\mathcal{S}}_1}
\bigg[\int d\pi\,Y^{-1}e^{i{\mathcal{S}}_2}(\hat{\unbu}_1+\hat{\unbu}_0\pi_1\pi_2)\bigg]
\end{aligned}
\end{equation}
where for the sake of notational simplicity, we put
$$
\begin{gathered}
{\mathcal{S}}={\mathcal{S}}_1+{\mathcal{S}}_2,\\
{\mathcal{S}}_1=\langle x|\xi\rangle+X\theta_1\theta_2,\quad
{\mathcal{S}}_2=Y\langle\theta|\pi\rangle+Z\pi_1\pi_2.
\end{gathered}
$$

Since
$$
\int d\pi\,Y^{-1}e^{i{\mathcal{S}}_2}(\hat{\unbu}_1+\hat{\unbu}_2\pi_1\pi_2)
=Y^{-1}\hat{\unbu}_0+Y^{-1}(iZ+Y^2\theta_1\theta_2)\hat{\unbu}_1,
$$
therefore, remarking $e^{iX\theta_1\theta_2}=1+iX\theta_1\theta_2$, we have
$$
\begin{aligned}
(1+iX\theta_1\theta_2)(Y^{-1}\hat{\unbu}_0&+Y^{-1}(iZ+Y^2\theta_1\theta_2)\hat{\unbu}_1)\\
&=Y^{-1}\hat{\unbu}_0+i Y^{-1}Z\hat{\unbu}_1
+[iXY^{-1}\hat{\unbu}_0-(Y+XY^{-1}Z)\hat{\unbu}_1]\theta_1\theta_2.
\end{aligned}
$$
Finally, we have
\begin{equation}
\begin{gathered}
({\mathcal{T}}_t\unbu)(\theta)=u_0(t,x)+u_1(t,x)\theta_1\theta_2  \with\\
{\begin{aligned}
u_0(t,x)&=(2\pi)^{-1/2}\int d\xi\, e^{i\langle x|\xi\rangle}(Y^{-1}\hat{\unbu}_0+iY^{-1}Z\hat{\unbu}_1),\\
u_1(t,x)&=(2\pi)^{-1/2}\int d\xi\, e^{i\langle x|\xi\rangle}[iXY^{-1}\hat{\unbu}_0-(Y+XY^{-1}Z)\hat{\unbu}_1].
\end{aligned}}
\end{gathered}
\end{equation}

Therefore, when ${\unbu}_1=0$, we have
\begin{equation}
u_0(t,x)=(2\pi)^{-1/2}\int d\xi \, e^{i\langle x|\xi\rangle} \varphi(t,x,\xi)\hat{\unbu}_0(\xi).
\label{chi-rv}
\end{equation}

\subsubsection{Calculation of $\varphi$}
We try to find a power series solution w.r.t. $t$ of \eqref{chi-pow}.
Putting\footnote{How to find factor $e^{it^2\xi/2}$? This is explained in Inoue [48].} $\varphi(t)=\psi(t)e^{it^2\xi/2}$,
$\psi$ satisfies below from \eqref{chi-pow}:
\begin{equation}
\ddot{\psi}+2it\xi\dot{\psi}+i(1-b)\xi\psi=0\with \psi(0)=1,\; \dot{\psi}(0)=0.
\label{chi-pow0}
\end{equation}

Putting $\alpha=2i\xi$, $\beta=i(1-b)\xi$, we rewrite \eqref{chi-pow0} as
\begin{equation}
\ddot{\psi}+\alpha t\dot{\psi}+\beta\psi=0\with \psi(0)=1,\; \dot{\psi}(0)=0.
\label{chi-pow1}
\end{equation}

Putting $\psi(t)=\sum_{j=0}^\infty c_jt^j$ into above, comparing coefficients of each $t^{\ell}$,
we have, for any $\ell$, $c_{2\ell+1}=0$ and
$$
c_{2\ell}=\frac{(-1)^{\ell}(2\alpha)^{\ell}}{(2\ell)!}\big(\frac{\beta}{2\alpha}\big)_{\ell}
\where (x)_{\ell}=x(x+1){\cdots}(x+\ell-1)=\frac{\Gamma(x+\ell)}{\Gamma(x)}.
$$

\begin{remark} (1) When $\beta=b-1=4k$ with $k=0,1,2,{\cdots}$, 
since
$$
\big(\frac{\beta}{2\alpha}\big)_{\ell}=(-k)_{\ell}=(-k)(-k+1){\cdots}(-k+\ell-1)=(-1)^\ell(k-1){\cdots}(k-\ell+1)
=\frac{(-1)^{\ell}k!}{(k-\ell)!}
$$
above power series becomes a finite sum, we have
\begin{equation}
\psi(t)=\sum_{j=0}^k \frac{4^jk!}{(2j)!(k-j)!}(i\xi)^jt^{2j}.
\label{conc-phi}
\end{equation}
(2) In case $b$ isn't represented as above, formal power series may diverge. Therefore, we need to impose conditions on initial data like Gevrey class etc.(see, \cite{ino24-1}) 
\end{remark}

At last, putting \eqref{conc-phi} into \eqref{chi-rv}, we get
$$
\begin{aligned}
u_0(t,q)&=(2\pi)^{-1/2}\int dp\, e^{iqp}\bigg(\sum_{j=0}^k \frac{4^jk!}{(2j)!(k-j)!}(ip)^jt^{2j}\bigg)e^{ip t^2/2}\hat{\unbu}_0(p)\\
&=(2\pi)^{-1/2}\int dp\, e^{ip(q+t^2/2)}\bigg(\sum_{j=0}^k \frac{2^{2j}k!}{(2j)!(k-j)!}(ip)^jt^{2j}\bigg)\hat{\unbu}_0(p)\\
&=\sum_{j=0}^k \frac{2^{2j}k!}{(2j)!(k-j)!}t^{2j}{\unbu}^{(j)}_0\bigg(q+\frac{t^2}{2}\bigg).
\end{aligned}
$$

\begin{remark}
(1) Since the coefficients of above $L(t,\partial_q)$ is $q$ independent, we may apply Fourier transformation w.r.t. $q$, getting
\begin{equation}
\hat{u}_{tt}(t,p)+t^2p^2\hat{u}(t,p)-ibp\hat{u}(t,p)=\hat{f}(t,p)\with \hat{u}(0,p)=\hat{\unbu}_0(p),   \hat{u}_t(0,p)=\hat{\unbu}_1(p).
\label{chiof}
\end{equation}
This operator is ``analogous" to \eqref{chi-pow}.\\
(2) Our procedure here may be adopted to the following Ivrii type operator:
\begin{equation}
L(t,q,\partial_q)=t^4\partial_{qq}-bq\partial_q.
\end{equation}
More precisely, see the forthcoming paper \cite{ino24-1}.
\end{remark}

\subsection{An application to Random Matrix Theory}
In 1983, Efetov \cite{efe83} wrote a paper entitled ``Supersymmetry and theory of disordered metals" where he derived Wigner's semi-circle law by applying ``his superanalysis".

Let ${\mathfrak{H}}_N$ be a set of $N\times N$ Hermite matrices. Identifying this topologically with ${\mathbb R}^{N^2}$, we  introduce probability measure $d\mu_N({H})$ on ${\mathfrak{H}}_N$.
\begin{equation}
\begin{gathered}
d\mu_N({H})=
\prod_{k=1}^N d(\Re H_{kk})\prod_{j<k}^N d(\Re H_{jk}) d(\Im H_{jk})
P_{N,J}(H),\\
P_{N,J}(H)=Z^{-1}_{N,J}\exp\big[-{\frac{N}{2J^2}}\tr {H}^*{H}\big].
\end{gathered}
\label{rmt-1}
\end{equation}
Here, ${H}=(H_{jk})$, 
$H^*=(H^*_{jk})=(\overline{H}_{kj})={}^t\overline{H}$,
$\prod_{k=1}^N d(\Re H_{kk})\prod_{j<k}^N d(\Re H_{jk}) d(\Im  H_{jk})$ is the Lebesgue measure  on ${\mathbb R}^{N^2}$, $Z^{-1}_{N,J}$ is the normalized constant given by $Z_{N,J}=2^{N/2}(J^2\pi/N)^{3N/2}$.

Let $E_\alpha=E_\alpha({H})$ ($\alpha=1,\cdots,N$) be real eigenvalues of a given matrix ${H}\in{{\mathfrak{H}}_N}$.
For Dirac's delta $\delta$, we put
\begin{equation}
\rho_N(\lambda)=\rho_N(\lambda;{H})=
N^{-1}\sum_{\alpha=1}^N \delta(\lambda-E_\alpha(H)),
\label{rmt-2}
\end{equation}
and for any function $f$ on ${\mathfrak{H}}_N$, we consider 
$$
\big\langle f\big\rangle_N=\big\langle f(\cdot)\big\rangle_N
=\int_{{\mathfrak{H}}_N}d\mu_N({H})\,f({H}).
$$
\begin{thm}[Wigner's semi-circle law]
\begin{equation}
\lim_{N\to\infty}\big\langle \rho_N(\lambda)\big\rangle_N=w_{sc}(\lambda)=
\begin{cases}
(2{\pi}J^2)^{-1}\sqrt{4J^2-\lambda^2}&\for |\lambda|<2J,\\
0 & \for |\lambda|>2J.
\end{cases}
\label{rmt-3}
\end{equation}
\end{thm}

This expression is derived by introducing auxiliary odd variables $\rho_1,\rho_2$:
\begin{equation}
\big\langle \rho_N(\lambda)\big\rangle_N=\pi^{-1}\Im
\int_{\mathfrak Q} dQ\,\big(\{(\lambda-i0){\mathbb{I}}_2-Q\}^{-1}\big)_{bb}
\exp{[-N{\mathcal I(q)}]}.
\label{rmt-4}
\end{equation}
Here $I_n$ is the $n\times n$ identity matrix, and
\begin{equation}
\begin{gathered}
{\mathcal I(q)}=\str[(2J^2)^{-1}Q^2+\log((\lambda-i0){\mathbb{I}}_2-Q)],\\
{\mathfrak Q}=\big\{Q=\begin{pmatrix}
x_1&\rho_1\\
\rho_2&ix_2
\end{pmatrix}\,\big|\, x_1, x_2\in{\rev},\; \rho_1,\rho_2\in{\rod}\big\}
\cong{\mathfrak R}^{2|2},\\
dQ=\frac{dx_1 dx_2}{2\pi} d\rho_1 d\rho_2,\\
\big(((\lambda-i0)I_2-Q)^{-1}\big)_{bb}
=\frac{(\lambda-i0-x_1)(\lambda-i0-ix_2)+\rho_1\rho_2}
{(\lambda-i0-x_1)^2(\lambda-i0-ix_2)}.
\end{gathered}
\label{rmt-5}
\end{equation}
For $2\times2$-supermatrix $A$, $(A)_{bb}$ is boson-boson(or even-even) part, in this case $A_{11}$.

Here, the key point of applying superanalysis is that in \eqref{rmt-4} parameter $N$ appeared in one place\footnote{This fact makes us remind the appearance of $\hbar$ at one place as in Feynman Path Integral representation},  and this derivation isn't justified mathematically there. In physics literature, they apply formally saddle point method or steepest descent to \eqref{rmt-4} when $N\to\infty$. Since they get
$$
\langle\frac{\delta{\mathcal{L}}(Q)}{\delta Q},\tilde{Q}\rangle
=\frac{d}{d\epsilon}{\mathcal{L}}(Q+\epsilon\tilde{Q})\bigg|_{\epsilon=0},
$$
critical point is given by 
$$
\frac{\delta{\mathcal{L}}(Q)}{\delta Q}=\str \big(\frac{Q}{J^2}-\frac{1}{\lambda-Q}\big)=0.
$$
Defining these as effective saddle points, putting 
$$
Q_c=\big(\frac12\lambda+\frac12\sqrt{\lambda^2-4J^2}\big){\mathbb{I}}_2,
$$
they have
$$
\lim_{N\to\infty}\big\langle \rho_N(\lambda)\big\rangle_N
=\pi^{-1}\Im (\lambda-Q_c)^{-1}_{bb}=w_{sc}(\lambda).  \qquad\qquad\square
$$
\begin{remark}
Though the derivation of expression \eqref{rmt-4} is justified in Inoue and Nomura \cite{IN00},
we can't justify the usage of saddle point method in this setting\footnote{Therefore, we give a proof of a very small portion which is explained in  \cite{fyo95}}, at that time.
Seemingly, this problem has relation to ``Laplace's method in function space" in \cite{sim79}. .
\end{remark}

\section{Some examples of FDE(=Functional Derivative Equation)}
\subsection{A simple Schwinger-Dyson equation as a FDE}
\subsubsection{A derivation of A Schwinger-Dyson equation of first order}
A formally given quadratic Lagrangian, we try to give meaning to the first order Schwinger-Dyson equation in Inoue \cite{ino86}: 
Let a Lagrangian be given through
\begin{equation}
\begin{aligned}
S(q,v)&=\frac12\int_{\euc} dt (\dot{q}(t)^2-\omega_0^2q(t)^2)\\
&\qquad+\frac12\int_{\euc^4}dxdt(|v_t(x,t)|^2-|\nabla v(x,t)|^2)
-\lambda\int_{\euc^4}dxdt\,\delta(x)q(t)v(x,t),
\end{aligned}
\label{SD1}
\end{equation}
then critical point of $S$ is calculated by
\begin{equation}
\begin{aligned}
\frac{d}{d\epsilon}S(q+\epsilon \rho,v)\bigg|_{\epsilon=0}
&=\int_{\euc} dt \bigg(-\frac{d^2}{dt^2}q(t)-\omega_0^2q(t)-\lambda\int_{\euc^3}dx\,\delta(x)v(x,t)\bigg)\rho(t)\\
&=\langle \frac{\delta S(q,v)}{\delta q(t)},\rho(t)\rangle=0,
\end{aligned}
\label{SD2-1}
\end{equation}
and
\begin{equation}
\begin{aligned}
\frac{d}{d\epsilon}S(q,v+\epsilon\varphi)\bigg|_{\epsilon=0}
&=-\int_{\euc^4}dxdt \,(v_{tt}(x,t)-\Delta v(x,t)+\lambda \delta(x)q(t))\varphi(x,t)\\
&=\langle \frac{\delta S(q,v)}{\delta v(x,t)},\varphi(x,t)\rangle=0.
\end{aligned}
\label{SD2-2}
\end{equation}
From these, at least formally, we may suppose the corresponding classical orbit $(q(t),v(x,t))$ satisfy below:
\begin{equation}
\left\{\begin{aligned}
\big(\frac{d^2}{dt^2}+\omega_0^2\big)q(t)&=-\lambda\int_{\euc^3}dx\,\delta(x)q(t)v(x,t),\\
\square v(x,t)&=-\lambda\delta(x)q(t).
\end{aligned}
\right.
\label{SDC1}
\end{equation}
Putting the generating functional as
\begin{equation}
\hat{Z}=\hat{Z}(q,v)=e^{i\hbar^{-1}S(q,v)},
\label{SD2-30}
\end{equation}
we define its characteristic functional,
\begin{equation}
Z(p,u)=Z_0^{-1}\int d_Fq \, d_Fv\, \hat{Z}(q,v)e^{-i\hbar^{-1}(\langle q,p\rangle+\langle v,u\rangle)}
\label{SD2-31}
\end{equation}
with
$$
Z(0,0)=Z_0=\int d_Fq \, d_Fv\, \hat{Z}(q,v), \quad \langle q,p\rangle=\int_{\euc}dt\,q(t)p(t), \quad
\langle v,u\rangle=\int_{\euc^4}dx dt \, u(x,t) v(x,t).
$$

This $Z(p,u)$ satisfies formally the following equation:
\begin{equation}
\left\{
\begin{aligned}
(\frac{d^2}{dt^2}+\omega_0^2)\frac{\delta Z(p,u)}{\delta p(t)}
&=i\hbar^{-1}p(t)Z(p,u)-\lambda\int_{\euc^3}dx\delta(x)\frac{\delta Z(p,u)}{\delta u(x,t)},\\
\square\frac{\delta Z(p,u)}{\delta u(x,t)}
&=i\hbar^{-1}u(x,t)Z(p,u)-\lambda\delta(x)\frac{\delta Z(p,u)}{\delta p(t)}.
\end{aligned}
\right.
\label{SD3}
\end{equation}

In fact, since
\begin{align}
\frac{\delta \hat{Z}(q,v)}{\delta q(t)}&=-i\hbar^{-1} \big(\frac{d^2}{dt^2}+\omega_0^2\big)q(t)\hat{Z}(q,v)-i\hbar^{-1}\lambda\int_{\euc^3}dx\delta(x)v(x,t) \hat{Z}(q,v), \label{5.71}\\
\frac{\delta \hat{Z}(q,v)}{\delta v(x,t)}&=-i\hbar^{-1}\square v(x,t)\hat{Z}(q,v)-i\hbar^{-1}\lambda\delta(x)q(t)\hat{Z}(q,v), \label{5.72}
\end{align}
assuming that above functional integral permit the integration by parts, we have
\begin{align}
&Z_0^{-1}\int d_Fq \, d_Fv\,\frac{\delta \hat{Z}(q,v)}{\delta q(t)}e^{-i\hbar^{-1}(\langle q,p\rangle+\langle v,u\rangle)}
=i\hbar^{-1}p(t)Z(p,u),\label{5.73}\\
&Z_0^{-1}\int d_Fq \, d_Fv\,\frac{\delta \hat{Z}(q,v)}{\delta  v(x,t)}e^{-i\hbar^{-1}(\langle q,p\rangle+\langle v,u\rangle)}
=i\hbar^{-1}u(x,t)Z(p,u).\label{5.74}
\end{align}
Interchanging integration and differentiation formally, we get
\begin{align}
&\frac{\delta Z(p,u)}{\delta p(t)}=-i\hbar^{-1}Z_0^{-1}\int d_Fq \, d_Fv\, q(t)\hat{Z}(q,v)e^{-i\hbar^{-1}(\langle q,p\rangle+\langle v,u\rangle)}, \label{5.75}\\
&\frac{\delta Z(p,u)}{\delta u(x,t)}=-i\hbar^{-1}Z_0^{-1}\int d_Fq \, d_Fv\, v(x,t)\hat{Z}(q,v)e^{-i\hbar^{-1}(\langle q,p\rangle+\langle v,u\rangle)}.\label{5.76}
\end{align}
Putting \eqref{5.71} into \eqref{5.73} and using \eqref{5.75} and \eqref{5.76}, we have
$$
\begin{aligned}
Z_0^{-1}\int d_Fq \, d_Fv\,&\bigg(-i\hbar^{-1} \big(\frac{d^2}{dt^2}+\omega_0^2\big)q(t)\hat{Z}-i\hbar^{-1}\lambda\int_{\euc^3}dx\delta(x)v(x,t) \hat{Z}\bigg)
e^{-i\hbar^{-1}(\langle q,p\rangle+\langle v,u\rangle)}\\
&= (\frac{d^2}{dt^2}+\omega_0^2)\frac{\delta Z(p,u)}{\delta p(t)}+\lambda\int_{\euc^3}dx\delta(x)\frac{\delta Z(p,u)}{\delta u(x,t)}=i\hbar^{-1}p(t)Z(p,u).
\end{aligned} 
$$
Analogously,
$$
\begin{aligned}
Z_0^{-1}\int d_Fq \, d_Fv\,&
\big(-i\hbar^{-1}\square v(x,t)\hat{Z}(q,v)-i\hbar^{-1}\lambda\delta(x)q(t)\hat{Z}(q,v)\big)
e^{-i\hbar^{-1}(\langle q,p\rangle+\langle v,u\rangle)}\\
&=\square\frac{\delta Z(p,u)}{\delta u(x,t)}+\lambda\delta(x)\frac{\delta Z(p,u)}{\delta p(t)}
=i\hbar^{-1}u(x,t)Z(p,u).
\qquad /\!/
\end{aligned} 
$$

\subsubsection{Reformulation and calculation}
\paragraph{Feynman propagator $E(x,t)=\square_F^{-1}(x,t)$.}
As a formal solution of division problem $\square E(x,t)=\delta(x,t)$ in $\mathcal{D}'(\euc^4)$,  we have
$$
E(x,t)=\lim_{\epsilon\to 0}(2\pi)^{-4}\int_{\euc^4}d\tau d\xi \frac{e^{-it\tau+ix\xi}}{-\tau^2+|\xi|^2-i\epsilon}
$$
whose precise meanings are given in, for example, Gelfand and Shilov \cite{GS64}. 
\begin{remark}
Approximate $\delta(x)$ by $\rho_\epsilon(x)=\epsilon^{-2}\rho(x/\epsilon)$, i.e. for $\rho(x)=\rho(|x|)\in C_0^{\infty}(\euc^3)$, $\rho(x)\ge0$  and $\int_{\euc^3}dx\rho(x)=1$,
making $\epsilon\to0$ then $\rho_\epsilon(x)\to\delta(x)$ in ${\mathcal{D}}'(\euc^3)$.
\end{remark}
\begin{lem}
For any $u\in{\mathcal{S}}(\euc^4)$, $\square_F^{-1}u\in{\mathcal{S}}'(\euc^4)$. Moreover,  when $u\in C_0^{\infty}(\euc^4)$,
$\lim_{\epsilon\to0}\langle \rho_\epsilon(\cdot), \square_F^{-1}u(\cdot,t)\rangle$ exists in $H^{-1}(\euc)$
and formally represented  as $\langle \delta, \square_F^{-1}u\rangle(t)$ or $\langle \delta(\cdot), \square_F^{-1}u(\cdot,t)\rangle$.
\end{lem}

\paragraph{\bf Renormalization.}
Applying $\square_F^{-1}$ to the second equation of \eqref{SD3}, and putting this into the first equation of \eqref{SD3}, we get
\begin{equation}
\begin{aligned}
(\frac{d^2}{dt^2}+\omega_0^2-\lambda^2&\int_{\euc^3}dx\,\delta(x)\square_F^{-1}(x,t))\frac{\delta Z(p,u)}{\delta p(t)}\\
&=-i\hbar^{-1}p(t)Z(p,u)+i\hbar^{-1}\lambda\int_{\euc^3}dx\,\delta(x)\square_F^{-1}u(x,t) Z(p,u).
\end{aligned}
\end{equation}
Here, we approximate $\delta(x)$ by $\rho_\epsilon(x)$. In this case, putting 
$$
(A_{\lambda}^\epsilon f)(t)=(\frac{d^2}{dt^2}+\omega_0^2)f(t)-\lambda^2\langle \rho_\epsilon(\cdot),(\square_F^{-1}(\rho_\epsilon f))(\cdot,t)\rangle,
$$ 
using Fourier transformation and Plancherel theorem,
$$
\begin{aligned}
(\widehat{A_{\lambda}^\epsilon f})(\tau)&=\frac{1}{\sqrt{2\pi}}\int d\tau e^{i t\tau}(A_{\lambda}^\epsilon f)(t)\\
&=(-\tau^2+\omega_0^2)\hat{f}(\tau)+\frac{\lambda^2}{4\pi^2}\int_{\euc^3}d\xi \frac{|\hat{\rho}_\epsilon(\xi)|^2}{\tau^2-|\xi|^2+i0} \hat{f}(\tau)\\
&=(-\tau^2+\omega_0^2)\hat{f}(\tau)\\
&\qquad\qquad+\frac{\lambda^2}{4\pi^2}\bigg[\int_{\euc^3}d\xi |\hat{\rho}_\epsilon(\xi)|^2
\bigg(\frac{1}{\tau^2-|\xi|^2+i0}+\frac{1}{|\xi|^2}\bigg)-\int_{\euc^3}d\xi \frac{|\hat{\rho}_\epsilon(\xi)|^2}{|\xi|^2}\bigg]\hat{f}(\tau).
\end{aligned}
$$
Simply disregarding the last term which is $\infty$, making $\epsilon\to0$, then we have 
$$
\begin{aligned}
\lim_{\epsilon\to0}\int_{\euc^3}d\xi |\hat{\rho}_\epsilon(\xi)|^2&\bigg(\frac{1}{\tau^2-|\xi|^2+i0}+\frac{1}{|\xi|^2}\bigg)\\
&=\int_{\euc^3}d\xi \bigg(\frac{\tau^2+i0}{(\tau^2-|\xi|^2+i0)|\xi|^2}\bigg)
=\int_{S^2}d\omega\int_0^{\infty}{\rho}^2d{\rho}\frac{\tau^2+i0}{(\tau^2-{\rho}^2+i0){\rho}^2}\\
&=4\pi\int_0^{\infty}d{\rho}\frac{\tau^2+i0}{\tau^2-{\rho}^2+i0}=-4\pi(2\pi i)\frac{\tau^2}{2|\tau|}=-4i\pi^2|\tau|
\end{aligned}
$$
Therefore
$$
\lim_{\epsilon\to0}(\widehat{A_{\lambda}^\epsilon f})(\tau) =\{(-\tau^2+\omega_0^2)-i{\lambda^2}|\tau|\}\hat{f}(\tau),
$$
so we define
$$
(A_{\lambda}^R f)(t)=\bigg(\frac{d^2}{dt^2}+\omega_0^2-i\lambda^2\big|\frac{d}{dt}\big|\bigg)f(t).
$$
When $\lambda>0$, for any $\tau\in\euc$, since $-\tau^2-{i\lambda^2}|\tau|+\omega_0^2\neq0$, we have
\begin{lem}
Operator $A_{\lambda}^R (\lambda>0)$ is invertible,  and $(A_{\lambda}^R)^{-1}$ maps $H^{-1}(\euc)$ to $H^1(\euc)$ as bounded operator.  Moreover, for any $p\in H^1(\euc)$,  the quantity $\langle (A_{\lambda}^R)^{-1}p,p\rangle$ is well-defined.
\end{lem}
\begin{problem}
Though I myself wrote this  paper, it isn't clear the sentence ``simply disregarding the last term". For example, whether may we find analogous regularization, such as zeta regularization for divergent series or divergent integral in Aghili and Tafazoli \cite{AT18}
or other summability methods?
\end{problem}
\paragraph{\bf Renormalized FDE equation and a result}
Now, we have the renormalized FDE equation
\begin{equation}
\begin{aligned}
A_{\lambda}^R\frac{\delta Z(p,u)}{\delta p(t)}&=-i\hbar^{-1}p(t)Z(p,u)+i\hbar^{-1}\lambda\int_{\euc^3}dx\delta(x)\square_F^{-1}u(x,t) Z(p,u),\\
\square \frac{\delta Z(p,u)}{\delta u(x,t)}&=[i\hbar^{-1}u(x,t)+i\hbar\lambda^2(\delta(x)(A_{\lambda}^R)^{-1}\langle\delta(\cdot),\square_F^{-1}u(\cdot,t)\rangle)\\
&\qquad\qquad\qquad\qquad\qquad\qquad-i\hbar^{-1}\lambda(\delta(A_{\lambda}^R)^{-1}p)(x,t)]Z(p,u)
\end{aligned}
\end{equation}
whose solution is obtained explicitly as
\begin{equation}
\begin{aligned}
Z(p,u)=\exp &
{[i(2\hbar)^{-1}\langle (A_{\lambda}^R)^{-1}p,p\rangle+i(2\hbar)^{-1}\langle \square_F^{-1}u,u\rangle}\\
&\qquad
{+i(2\hbar)^{-1}\lambda^2\langle(A_{\lambda}^R)^{-1}\langle\delta,\square_F^{-1}u\rangle(t),\langle\delta,\square_F^{-1}u\rangle(t)\rangle}\\
&\qquad\qquad
{-i\hbar^{-1}\lambda\langle (A_{\lambda}^R)^{-1}p,\langle\delta,\square_F^{-1}u\rangle(t)\rangle]}
\end{aligned}
\end{equation}
and  for any $n, m\in {\mathbb{N}}$, $\{t_j\}_{j=1}^n\subset \euc$, $\{(y_k,s_k)\}\subset \euc^3\times\euc$, all Green function are obtained as bellow:
$$
\begin{aligned}
G^{(n,m)}&(t_1,\cdots, t_n,(y_1,s_1),\cdots,(y_m,s_m)))\\
&\qquad
=(i\hbar)^{n+m}\frac{\delta^{n+m}Z(0,0)}{\delta p(t_1)\cdots\delta p(t_n)\delta u(y_1,s_1)\cdots\delta u(y_m,s_m)}.
\end{aligned}
$$

\subsubsection{An extension of Lagrangian~\eqref{SD1} and a problem}
Author learned the Lagrangian \eqref{SD1} from Arai \cite{ara81} 
which is originally studied by Aichelburg and Grosse \cite{AG77} 
as
\begin{equation}
\begin{aligned}
I(q_1,q_2,v)&=\frac{1}{2}\int_{\euc}dt(\dot{q}_1(t)^2-\omega_0^2 q_1(t)^2+
\dot{q}_2(t)^2-\omega_0^2 q_2(t)^2)\\
&\qquad+\frac{1}{2}\int_{\euc^4}dxdt(|v_t(x,t)|^2-|\nabla v(x,t)|^2)\\
&\qquad\qquad
-\lambda\int_{\euc^4}dxdt(\delta(x-a)q_1(t)+\delta(x+a)q_2(t))u(x,t).
\end{aligned}
\label{AG}
\end{equation} 
\begin{problem}  Find the renormalized Schwinger-Dyson equation corresponding to \eqref{AG} and write down the Green functions as before! Increase the number of points where we put harmonic oscillators, and obtain Green functions! Moreover, what occurs when we put randomly these oscillators and finally consider as Ozawa did in \cite{oza90}? 
\end{problem}
\par
{\bf{Small goals}}: Enlarging Feynman's idea slightly, following afore mentioned Itzykson-Zuber's claim,
we dare to imagine
``{{Quantum mechanics(Quantum field theory) stands for obtaining properties of generating function\footnote{For the time being, we don't concern with how these functions relate to the physical quantities observed by experiments}
represented by Lagrange or Hamilton function using Feynman measure}''. To relate these quantities to
classical mechanics(classical field theory)
and to get rid of the use of non-existing Feynman measure, we take formally the characteristic function which satisfies certain FDE\footnote{Since in general, it contains higher order derivatives, we need new device contrivance such as Colombeau's generalized functions, see for example Gsponer \cite{gsp06}, at least in there, we may multiply two generalized functions freely though not sure taking the trace of diagonal part is permissible} and give meaning to it after renormalization and solve it!

As is known, to study PDE without getting explicit solution of it, we fully use Lebesgue measure to have the existence proof using integration by parts, change of variables formula under integral sign and Fourier transformations with the method of functional analysis\footnote{Project  called  ``A study of PDE by functional analytic method'' leaded by K\^osaku Yosida in Japan.}. As mentioned before, not only there doesn't exist a translational invariant, completely additive measure in function spaces, but also giving meaning to the trace of higher order functional derivatives is hard.
For a simple quadratic interaction, we give meaning to some devices eliminating $\infty$. This suggests that as is used by physicists, expand w.r.t. interaction parameter and performing Gaussian type integration, after summation method, we get something-like a solution of FDE. 

Moreover, in the next subsection, I give a trial to understand the trace of 2nd order functional derivatives following Inoue \cite{ino87}. 

\subsection{Hopf equation as a FDE}
\paragraph{\bf Differential geometric preliminaries and notations}
Let $(M, g_{jk})$ be a compact Riemannian manifold of dimension $d$ with or without boundary $\partial M$.
We denote by ${\overset{\circ}{\mathbf{X}}}_{\sigma}(M)$ and ${\overset{\circ}{\Lambda}}{}_{\sigma}^1(M)$, the space of all solenoidal vector fields on $M$ which vanish near the boundary and that of all divergence free $1$-forms on $M$ which vanish near the boundary, respectively. $\tilde{\mathbf{H}}$ (resp. $\mathbf{H}$) stands for the completion of the space ${\overset{\circ}{\Lambda}}{}_{\sigma}^1(M)$ (resp. ${\overset{\circ}{\mathbf{X}}}_{\sigma}(M)$) w.r.t. $\tilde{\mathbf{L}}^2$-norm (resp. ${\mathbf{L}}^2$-norm).

We denote by $\tilde{\bf L}^2, \tilde{\bf H}, \tilde{\bf H}^s, \tilde{\bf V}, \tilde{\bf W}$ and $\tilde{\bf V}^s$, the completion of $\Lambda^1(\bar{M})$ or $\Lambda_{\sigma}^1(\bar{M})$ or ${\overset{\circ}{\Lambda}}{}_{\sigma}^1(M)$ with respect to the corresponding norms which are represented by the same symbols.

The dual space of $\bf H$ w.r.t. $\langle{\cdot}, {\cdots}\rangle$ or $({\cdot}, {\cdots})$ is denoted by $\tilde{\bf H}$ or $\bf H$, respectively. Analogously, the dual space of ${\bf V}^s$ w.r.t. $\langle{\cdot}, {\cdots}\rangle$ or $({\cdot}, {\cdots})$ is denoted by $\tilde{\bf V}^{-s}$ or ${\bf V}^{-s}$, respectively. 

The space of symmetric  tensor fields with $k$ contravariant (resp. covariant) indeces is denoted by $ST_k(M)$(resp. $ST^k(M))$. For example $w\in ST_2(M)$ means that it is expressed locally as
$$
w=w^{ij}\frac{\partial}{\partial x^i}\otimes\frac{\partial}{\partial x^j}
$$
with some function $w^{ij}=w^{ij}(x)$ on $M$, symmetric in $i, j$. Analogously, $\zeta\in ST^2(M)$ stands for
$$
\zeta=\zeta_{ij}\, dx^i\otimes dx^j
$$ 
with some functions $\zeta_{ij}=\zeta_{ij}(x)$ locally on $M$, symmetric in $i, j$.
 
[notation]:
$$
\begin{aligned}
\Gamma^{\ell}_{ij}&=\frac{1}{2}g^{\ell k}\{\frac{\partial}{\partial x^i}g_{kj}+ \frac{\partial}{\partial x^j}g_{ki}- \frac{\partial}{\partial x^k}g_{ij}\},\\
R^{\ell}_{ijk}&=\frac{\partial}{\partial x^k}\Gamma^{\ell}_{ij} - \frac{\partial}{\partial x^j}\Gamma^{\ell}_{ik}+\Gamma^m_{ij}\Gamma^{\ell}_{mk}-\Gamma^m_{ik}\Gamma^{\ell}_{mj},\\
R_{ij}&=R^m_{ijm} \et R^i_j=g^{ik}R_{kj}
\end{aligned}
$$

For a $1$-form $\eta=\eta_j dx^j$, we define
$$
\begin{gathered}
(\nabla_i\eta)_j=\frac{\partial}{\partial x^i}\eta_j-\Gamma^{\ell}_{ij}\eta_{\ell},\\
(\nabla^i\eta)_j=g^{ik}(\nabla_k\eta)_j=g^{ik}\{\frac{\partial}{\partial x^k}\eta_j-\Gamma^{\ell}_{jk}\eta_{\ell}\},\\
\delta\eta=\frac{1}{\sqrt{g}}\frac{\partial}{\partial x^i}\{{\sqrt{g}}g^{ij}\eta_j\} \et (\Delta \eta)_j=\nabla^k\nabla_k\eta_j+R^k_j\eta_k.
\end{gathered}
$$

We identify a $1$-form $\eta$ with a vector field $v$ by
$$
v^i=g^{ij}\eta_j \et \eta_i=g_{ij}v^j,
$$
which expresses with abuse of notation, as
$$
v^i=(\tilde{\eta})^i \et \eta_i=(\tilde{v})_i.
$$

For a vector field $\displaystyle{u=u^j\frac{\partial}{\partial x^j}}$, we put
$$
\begin{gathered}
(\nabla_i u)^j=\frac{\partial}{\partial x^i}u^j+\Gamma^j_{ik}u^k,\\
(\nabla^i u)^j=g^{ik}(\nabla_ku)^j=g^{ik}\{\frac{\partial}{\partial x^k}u^j+\Gamma^j_{k\ell}u^{\ell}\},\\
\delta u=\frac{1}{\sqrt{g}}\frac{\partial}{\partial x^i}\{\sqrt{g}u^i\} \et (\Delta u)^j=\nabla^k\nabla_k u^j-R^j_k u^k.
\end{gathered}
$$

$$
\langle u,\eta\rangle=\int_Mu_i\eta^i d_gx \for u\in{\mathbf{X}}(\bar{M}) \et \eta\in \Lambda^1(\bar{M}),
$$

For $u, v\in{\mathbf{X}}(\bar{M}) $, we put
$$
(u,v)=\int _Mg_{ij}u^iv^jd_gx=\langle u,\tilde{v}\rangle=\langle \tilde{u},v\rangle \et (u,u)=|u|^2.
$$
For $\xi, \eta\in \Lambda^1(\bar{M})$, we put
$$
(\eta,\xi)=\int_M g^{ij}\eta_i\xi_j d_gx=\langle\eta,\tilde{\xi}\rangle=\langle\tilde{\eta},{\xi}\rangle \et (\eta,\eta)=|\eta|^2.
$$
$$
(\!(u,v))\!)=(\nabla_k u, \nabla^k v)=\int_M g_{ij}\nabla _ku^i\nabla^k v^j d_gx \et (\!(u,u))\!)=\Vert u\Vert^2.
$$

\subsubsection{Differential geometrical expression of Navier-Stokes equation}
For a vector field $u=u^j\displaystyle{\frac{\partial}{\partial x^j}}$, 
we have the Navier-Stokes equation
\begin{equation}
\frac{\partial u^i}{\partial t}-\nu(\Delta u)^i+(\nabla_u u)^i+p^i=f^i,
\label{NSvec}
\end{equation}
with
$$
\delta u=0, \quad u(0,x)={\unbu}(x) \et u(t,x)\big|_{x\in\partial M}=0.
$$

\subsubsection{Hopf's motivation}
Following Hopf's introduction in \cite{hop52}, we quote  
\begin{quotation}
The differential law which governs the motion of a deterministic mechanical system has the symbolic form
$$
\frac{du}{dt}={\mathfrak{F}}(u)
$$
where $u$ is an instantaneous phase of the system and where the righthand and side is completely determined by the phase.
$\cdots\cdots$
\par
There are mechanical systems the phases of which are characterized by a very large number of independent parameters and the phase motions of which are tremendously complicated.
Two examples are the classical model of a gas with its very large number of degrees of freedom and the flow of viscous incompressible fluid at a very large value of overall Reynolds number.
\par
In both cases the important task is not the determination of the exact phase motion with an exactly given initial phases but the determination of the statistical properties of the ``typical" phase motion.
In order to achieve this goal statistical mechanics studies probability distributions of simultaneous phases and their evolution in time resulting from the individual phase motions.
$\cdots\cdots$
\end{quotation}
Moreover, he claims to find statistical equilibrium of the system. 
This idea is comparable to find equilibrium state of Hamilton flow by finding a stable solution of Liouville equation.
Any way, Fois\c{c} considers this problem in \cite{foi73} 
but he doesn't regard this problem from the point of FDE.

\subsubsection{Derivation of Hopf equation}

Let $T_tu_0$ be a (some kind of) solution of Navier-Stokes equation~\eqref{NSvec} with the initial data $u_0$,
for a Borel measure $\mu(\omega)$ on ${\bf H}$, we put $\mu_t(\omega)=\mu(T_t^{-1}(\omega))$. Here, $\omega$ is a Borel set on $\bf{H}$. The characteristic function of this measure $\mu_t(\omega)$ should satisfy the equation, called Hopf equation. More precisely, 

(I) the characteristic function of this measure $d\mu_t(\cdot)$ is given by
$$
W(t,\eta)=\int_{\bf{H}}d\mu_t(u)\,e^{i\langle u,\eta\rangle} =\int_{\bf{H}}d\mu(u)\,e^{i\langle T_tu,\eta\rangle} 
$$
which satisfies
\begin{equation}
\begin{aligned}
\pdt W(t,\eta)
&=\int_M d_gx \left[ 
-i\{\frac{\partial}{\partial x^k}\eta_j(x)-\Gamma^{\ell}_{jk}(x)\eta_{\ell}\}
\frac{\delta^2 W(t,\eta)}{\delta\eta_j(x)\delta\eta_k(x)}\right.\\
&\qquad\qquad\qquad\qquad
\left.
+\nu(\Delta \eta)_j(x)\frac{\delta W(t,\eta)}{\delta\eta_j(x)}
+i\eta_j(x)f^j(t,x)W(t,\eta)\right].
\end{aligned}
\label{Hopf1}
\end{equation}

One of the additional conditions to this equation is
\begin{equation}
\frac{1}{\sqrt{g(x)}}\frac{\partial}{\partial x^j}\{\sqrt{g(x)}\frac{\delta W(t,\eta)}{\delta\eta_j(x)}\}=0,
\label{Hopf2}
\end{equation}
\begin{equation}
W(0,\eta)=W_0(\eta) \et W(t,0)=1.
\label{Hopf3}
\end{equation}
Here, $t\in(0,\infty)$ and
$$
\eta=\eta(x)=\eta_j(x)dx^j\in \overset{\circ}{\Lambda}{}^1_{\sigma}(M),\quad
f=f(t,x)=f^j(t,x)\partial/\partial x^j\in \overset{\circ}{\bf{X}}{}_{\sigma}(M).
$$
Given positive definite functional $W_0(\eta)$ on $\tilde{\bf H}$ satisfies
\begin{equation}
W_0(\eta)=1\et \frac{1}{\sqrt{g(x)}}\frac{\partial}{\partial x^j}\bigg[\sqrt{g(x)}\frac{\delta W_0(\eta)}{\delta\eta_j(x)}\bigg]=0.
\label{Hopf4}
\end{equation}
This equation is something like Liouville equation corresponding to Navier-Stokes equation.
In fact, let $u(t,x)=T_t{\unbu}(x)$ be a ``solution" for Navier-Stokes equation with initial data $\underline{u}=\underline{u}^j\partial/\partial x^j\in \mathbf{H}$. For Borel measure $\mu(\omega)$ on ${\bf H}=L^2_{\sigma}(\Omega)$, put $\mu_t(\omega)=\mu(T_t^{-1}(\omega))$ for any Borel set $\omega$ in $\bf{H}$.
More explicitly, put
\begin{equation}
\Phi(t,v)=\int \mu(du)\,e^{i(v, T_tu)}=\int \mu_t(du)\,e^{i(v,u)},
\label{hopf2}
\end{equation}
Then, this functional satisfies Hopf equation formally.
In the above, for $u$, we put $\tilde{u}$ as the solenoidal part of $u$.

\subsubsection{How to give the meaning to the trace of 2nd order functional derivative}
\begin{problem}
Our problems here are, whether does there exist a functional $W(t,\eta)$ satisfying \eqref{Hopf1}, and how to give meaning to the 2nd order functional derivatives $\displaystyle{\frac{\delta^2 W(t,\eta)}{\delta\eta_j(x)\delta\eta_k(x)}}$?
\end{problem}

Before proceeding further, we need some explanation to functional derivatives in general setting.
\begin{defn}
Let $\Phi=\Phi(f)$ be a functional $E=E(M)$, a suitable function space on $M$. If at $f\in E$, there exists $D\Phi(f)\in E'$ such that
$$
\frac{d}{d\epsilon} \Phi(f+\epsilon h)\big|_{\epsilon=0}=\langle D\Phi(f), h\rangle
=\int_M\frac{\delta \Phi(f)}{\delta f(x)}h(x)d_gx \for h\in C_0^{\infty}(M)
$$
then $\Phi$ is said to be (Gateaux)-differentiable at $f$ in the direction $h$ and
formally the rightest integral representation.
Or, though $\delta(\cdot)$ is not contained $E$, we may denote
$$
\frac{d}{d\epsilon} \Phi(f(\cdot)+\epsilon \delta(\cdot))\big|_{\epsilon=0}.
$$
\end{defn}
\begin{defn}
Let $\Phi=\Phi(f)$ be a differentiable functional on $E$. If $\langle D\Phi(f), h_1\rangle$ is differentiable as a functional of $f$ at each $h_1\in E$, then $\Phi$ is called twice differentiable at $f$ and its second derivative $D^2\phi(f)$ is given by
$$
\begin{aligned}
\frac{d}{d\epsilon}\langle D\Phi(f+\epsilon h_2), h_2\rangle\big|_{\epsilon=0}&=\langle D^2\Phi(f), h_1\otimes h_1\rangle\\
&=\int_{M\times M}\frac{\delta^2\Phi(f)}{\delta f(x) \delta f(y)} h_1(x) h_2(y) d_gx d_gy\\
&=\int_{M\times M}\frac{\delta^2\Phi(f)}{\delta f(y) \delta f(x)} h_1(x) h_2(y) d_gx d_gy.
\end{aligned}
$$
In right-hand side of the last equality above, $\langle{\cdot},{\cdot}\rangle$ stands for the duality between $E(M\times M)\sim E(M)\otimes E(M)$ and $E'(M\times M)\sim E'(M)\otimes E'(M)$ and $(h_1\otimes h_2)(x,y)=h_1(x)h_2(y)$, $h_1, h_2\in C_0^{\infty}(M)\subset E$ and $ x,y\in M$.
Since $\displaystyle{\frac{\delta^2\Phi(f)}{\delta f(x) \delta f(y)}}$ is a distribution on $M\times M$, applying Schwartz kernel theorem \cite{sch66}, 
we have formally the above. It generally does not make sense to put $x=y$. But if it is possible to denote it by $\displaystyle{\frac{\delta^2\Phi(f)}{\delta f(x)^2}}$ and call it the trace of $\displaystyle{\frac{\delta^2\Phi(f)}{\delta f(x) \delta f(y)}}$.
Higher order functional derivatives are defined analogously.
\end{defn}

Moreover, we define
\begin{defn}[Definition 2.12 in \cite{ino87}] 
(1) A functional $W$ on $\tilde{\bf H}$ is said to be positive definite when for any $n\in{\mathbb{N}}$, $v$ and $\{\eta_j\}_{j=1}^n$ in $\tilde{\bf H}$ if for any $\{\zeta_j\}_{j=1}^n\subset{\mathbb{C}}$ we have
$$
\sum_{k,j=1}^n \bar\zeta_k\zeta_j W(\eta_k-\eta_j)\ge 0.
$$
(2) A positive definite functional $W$ on $\tilde{\bf H}$ is said to be of $\tilde{\bf V}^{-\ell}$-exponential type for any ${\eta}\in \tilde{\bf H}$, if the function $s\to W(s\eta)$ defined on $\euc$ can be extended analytically to an entire function $W(\zeta:\eta)$ on the complex plane ${\mathbb{C}}$ satisfying
$$
|W(\zeta:\eta)|\le c_5{\cdot}e^{c_6|\Im\zeta \Vert \eta\Vert_{-s}}\quad \forall \zeta\in {\mathbb{C}}, \eta\in \tilde{\bf H}.
$$
where $c_5$ and $c_6$ are constants depending on functional $W$.
\end{defn}

Above problem (I) is restated as

(II)  Find a Borel measure $\mu(t,\cdot)$ on ${\bf H}$ which satisfies
\begin{equation}
\begin{aligned}
-\int_0^{\infty}\int_{\bf H}&d\mu(t,u)\frac{\partial \Phi(t,u)}{\partial t}-\int_{\bf H}d\mu_0(u)\Phi(0,u)\\
&=\int_0^{\infty}\int_{\bf H}\int_M d_gx d\mu(t,u)dt \left[-u^k(x)u^j(x)\frac{\partial}{\partial x^k}\frac{\delta\Phi(t,u)}{\delta u^j(x)}
+\Gamma^j_{k\ell}(x)u^k(x)u^\ell(x)\frac{\delta\Phi(t,u)}{\delta u^j(x)}\right.\\
&\qquad\qquad\qquad\qquad\qquad\qquad\qquad\qquad\qquad      
\left. +\nu\nabla_ku^j(x){\cdot}\nabla^k\frac{\delta\Phi(t,u)}{\delta u^j(x)}-f^j(t,x)\frac{\delta\Phi(t,u)}{\delta u^j(x)}\right]
\end{aligned}
\label{HopfII-1}
\end{equation}
for any suitable test functionals $\Phi(t,u)$.

\begin{remark}
(1) There appeared only 1st order Functional Derivatives  in  (II)!\\
(2) How to relate problems (I) and (ii) is explained in \cite{ino87} precisely.
\end{remark}

\begin{thm}[A part of Theorem $\rm A^{\prime}$ of \cite{ino87}] 
Let $f(\cdot)\in L^2((0,\infty);{\bf{V}}^{-1})$ be given and suppose a Borel measure $\mu_0$ on $\bf{H}$ satisfies
$$
\int_{\bf{H}}d\mu_0(u)(1+|u|^2)<\infty.
$$
Then, there exists a basic family $\{\mu(t,\cdot)\}_{0<t<\infty}$ of Borel measures on $\bf{H}$ such that
$$
\begin{aligned}
\int_0^{\infty}dt\,\int_{\bf{H}}d\mu(t,u)&\Phi_t(t,u)+\int_{\bf{H}}d\mu_0(u)\Phi_t(0,u)\\
&=\int_0^{\infty}dt\left[\int_{\bf{H}}d\mu(t,u)\left\{\nu a(u,\tilde{\Phi}_u(t,u))+b(u,u, \tilde{\Phi}_u(t,u)) 
-\langle f(t),{\Phi}_u(t,u)\rangle \right\}  \right]
\end{aligned}
$$
for any $\Phi\in TF$ with compact support in $t$, i.e. there exists a constant $T$ depending on $\Phi$ such that $\phi(t,\cdot)=0$ when $t\ge T$.
\end{thm}

In the above, forms $a(\cdot,\cdot)$ and $b(\cdot,\cdot,\cdot)$ are defined by
$$
\begin{gathered}
a(u,v)=\int_M d_gx\, g_{ij}\nabla_k u^i \nabla^kv^j,\\
b(u,v,w)=(\nabla_uv,w)=\int_M d_gx\, g_{ij}\{u^k\frac{\partial}{\partial x^k}v^i+\Gamma^i_{k\ell}u^kv^{\ell}\}w^j
\end{gathered}
$$
for $u,v,w\in{\bf{X}}(M)$ with $u=u^j\frac{\partial}{\partial x^j}$ etc.

\begin{defn}[Definition 2.7 in \cite{ino87}]
A real functional $\Phi(\cdot,\cdot)$ on $[0,\infty)\times \bf{V}$ is called TF(=test functionl) denoted by $\Phi\in TF$
if it satisfies the following.
\begin{enumerate}
\item $\Phi(\cdot,\cdot)$ is continuous on $[0,\infty)\times \bf{V}$ and verifies
$$
|\Phi_u(t,u)|\le c  \et  |\Phi_t(t,u)|\le c+c|u|
$$
where $\Phi_u(t,\cdot)$ is regarded as an element in $\tilde{\bf{H}}$.\\
\item $\Phi(\cdot,\cdot)$ is Fr\'echet $\bf{H}$-differentiable in the direction $\bf{V}$.\\
\item $\Phi_u(\cdot,\cdot)$ is continuous on $[0,\infty)\times \bf{V}$ to $\tilde{\bf{V}}^s$ and is bounded, i.e. 
there exists a constant $c$ depending on $\Phi$ such that
$$
\Vert \Phi_u(t,u)\Vert_s\le c \et |\Phi_t(t,u)|\le c+c|u|\quad  \forany (t,u)\in[0,\infty)\times \bf{V}.
$$
\end{enumerate}
\end{defn}
\begin{defn}[Definition 2.8 in \cite{ino87}]
A family of Borel measures on ${\bf{H}}$ is called s solution of Problem (II) on $(0,\infty)$ if it satisfies (II) and the following conditions:
\begin{enumerate}
\item 
$\displaystyle{\int_{\bf{H}}(1+|u|^2)d\mu({\cdot},u)\in L^{\infty}(0,\infty)}$,
\item 
$\displaystyle{\int_{\bf{H}}\Vert u\Vert^2 d\mu({\cdot},u)\in L^1(0,\infty)}$,
\item 
$\displaystyle{\int_{\bf{H}}\Phi(u) d\mu({\cdot},u)}$ is measurable in $t$ for any non-negative, weakly continuous functional $\Phi(\cdot)$ on $\bf{H}$,
\end{enumerate}
\end{defn}
\begin{defn}[Definition 2.9 in \cite{ino87}]
A functional defined on $[0,T)\times\tilde{\bf{H}}$, ($T\le{\infty}$) will be called a strong solution of Problem (I) on $(0,T)$ if there exists a set $\tilde{D}$, dense in $\tilde{\bf{V}}^s$, for some $s$, containing $\overset{\circ}{\Lambda}{}^1_{\sigma}(M)$ such that:
\begin{enumerate}
\item
For each $\eta\in \tilde{D}$, $W(t,\eta)$ belongs to $L^1_{loc}[0,T)$ is continuous in $t$ at $t=0$ and is twice differentiable at $\eta\in \tilde{D}$ for $a.e.t$.
\item
$\displaystyle{\frac{\delta^2 W(t,\eta)}{\delta \eta_j(x)\delta \eta_k(x)}\frac{\partial}{\partial x^j}\otimes\frac{\partial}{\partial x^k}}$
exists for almost every $t$ on $(0,T)$ as a distributional element in $ST_2(M)$ for $\eta\in \tilde{D}$.
\end{enumerate}
\end{defn}
\begin{defn}[Definition 2.10 in \cite{ino87}]
A functional defined on $[0,T)\times\tilde{\bf{H}}$, ($T\le{\infty}$) will be called a classical solution of Problem (I) on $(0,T9$ if there exists a dense set $\tilde{D}$ in $\tilde{\bf V}^s$ for some $s$, containing $\overset{\circ}{\Lambda}{}^1_{\sigma}(M)$ such that:
\begin{enumerate}
\item
$W(t,\eta)$ is absolutely continuous on $[0,T)$ for each $\eta\in \tilde{D}$ and is twice differentiable; moreover, $\displaystyle{\frac{\delta W(t,\eta)}{\delta \eta_j(x)}}$ belongs to $L^1_{loc}(M)$ for each $j$.
\item
$\displaystyle{\frac{\delta^2 W(t,\eta)}{\delta \eta_j(x)\delta \eta_k(x)}}$ exists for each $j, k$ and almost every $t$ on $[0,T9$ as an element of $L^1_{loc}(M)$ for $\eta\in \tilde{D}$. Moreover
$$
\frac{\delta^2 W(t,\eta)}{\delta \eta_j(x)\delta \eta_k(x)}\frac{\partial}{\partial x^j}\otimes\frac{\partial}{\partial x^k}
$$
belongs to $ST_2(M)$ as an elements of $L^1_{loc}(M)$.
\item
$W(t,\eta)$ satisfies (I.1)--(I.4) for almost every $t$ as functions for each $\eta\in\tilde{D}$.
\end{enumerate}
\end{defn}

\begin{thm}[Theorem A in \cite{ino87}]
Let $f(\cdot)\in L^2((0,\infty);{\bf{V}}^{-1})$ be given and suppose a 
For any Borel measure $\mu_0$ on $\bf{H}$ satisfying
$$
\int_{\bf{H}}d\mu_0(u)(1+|u|^2)<\infty,
$$
and any $f(\cdot)\in L^2((0,\infty);{\bf{V}}^{-1})$, there exists a solution $\{\mu(t,\cdot)\}_{0<t<\infty}$ of Problem (II).
\par
Moreover, it satisfies the following energy inequality of strong form.
$$
\begin{aligned}
\frac{1}{2}\int_0^{\infty}dt\,\int_{\bf{H}}d\mu(t,u)&\psi(|u|^2)+\nu\int_0^t\big[\int_{\bf{H}}d\mu(\tau,u)\psi'(|u|^2)\Vert u\Vert^2\big]d\tau\\
&\le\frac{1}{2}\int_{\bf{H}}d\mu_0(u)\psi(|u|^2)+\int_0^{\infty}d\tau\big[\int_{\bf{H}}d\mu(\tau,u)\psi'(|u|^2)f(\tau,u)\big]
\end{aligned}
$$
for $0<t<\infty$ and $\psi\in C^1(0,[\infty)$ satisfying
$$
0\le \psi'(t)\le \sup_{s\in[0,\infty)}\psi'(s)<\infty.
$$
\end{thm}

\begin{thm}[Theorem B in \cite{ino87}]
Suppose a positive definite functional $W_0({\cdot})$ on $\tilde{\bf H}$ satisfies
$$
\tr_{\tilde{\bf H}\to{\bf H}}[-W_{0\eta\eta}]<\infty.
$$
For any $f(\cdot)\in L^2((0,\infty);{\bf{V}}^{-1})$,  there exists a strong solution $W(t,\eta)$ of Problem (I).
\end{thm}
\begin{thm}[Theorem C in \cite{ino87}]
Let $\partial M=\emptyset$ and let $\ell$ be the largest integer not exceeding $(d/2)+1$. Suppose a positive definite functional $W_0(\cdot)$ on $\tilde{\bf H}$ is of $\tilde{\bf V}^{-\ell}-$ exponential type and satisfies
$$
\tr_{\tilde{\bf H}\to{\bf H}}[-W_{0\eta\eta}(0)]<\infty \et \tr_{\tilde{{\bf V}}^{\ell}\to{\bf V}^{\ell}}[-W_{0\eta\eta}(0)]<\infty.
$$
For any $f({\cdot})\in L^1_{loc}(0,{\infty}; {\bf V}^{\ell})$, there exists a classical solution $W(t,\eta)$ of Problem (I) on $[0,T^*)$ where $T^*<\infty$ is determined by $W_0$ and $f$ independently of $\nu$.
\end{thm}

\subsection{A Schwinger-Dyson equation corresponding to $P(u)^4$ model}

As a simple example of quantum field theory, we consider the so-called $P(u)^4$ model.

Consider the following functional
\begin{equation}
S(u)=\int dxdt \bigg(\frac{1}{2}\partial_t u^2-\frac{1}{2}|\nabla u|^2-\frac{\lambda}{4}u^4\bigg).
\end{equation}
The critical point of this functional should satisfy
\begin{equation}
\begin{aligned}
\langle\frac{\delta S(u)}{\delta u(t,x)},v(x,t)\rangle&=\frac{d}{d\epsilon}S(u+\epsilon v)\big|_{\epsilon=0}\\
&=
\int d^3xdt (u_t(t,x)v_t(t,x)-\nabla u(t,x)\nabla v(t,x)-\lambda u^3(x,t)v(x,t))\\
&=\langle -\square u(x,t)-\lambda u^3(x,t), v(t,x)\rangle.
\end{aligned}
\end{equation}
This is formally represented by
\begin{equation}
\frac{\delta S(u)}{\delta u(t,x)}=-\square u(x,t)-\lambda u^3(x,t).
\label{NLW}
\end{equation}

\begin{remark}
Concerning the initial value problem for semi-linear wave equation \eqref{NLW}, consult, for example Lai and Zhou \cite{LZ14} 
and Liu and Wang \cite{LW22}. 
\end{remark}

Quantum field theory seems to study the property of a formal Feynman measure
\begin{equation}
d_F u\, e^{i\hbar^{-1}S(u)}.
\end{equation}
To do this, we consider the characteristic functional of that measure;
\begin{equation}
\begin{gathered}
Z(\eta)=Z_0^{-1}\int d_F u\, e^{i\hbar^{-1}S(u)}e^{i\hbar^{-1}\langle u,\eta \rangle}\\
 \where \langle u,\eta \rangle=\int dx dt\, u(t,x)\eta(t,x)\et  Z_0=\int d_F u\, e^{i\hbar^{-1}S(u)},
\end{gathered}
\end{equation}
and its property should be studied by deriving the Schwinger-Dyson equation corresponding to above:
$$
\begin{aligned}
\frac{d}{d\epsilon} Z(\eta+\epsilon \phi)\big|_{\epsilon=0}
&=\frac{d}{d\epsilon}\int d_F u\, e^{i\hbar^{-1}S(u)}e^{i\hbar^{-1}\langle u,\eta+\epsilon \phi \rangle}\big|_{\epsilon=0}\\
&=\langle \frac{\delta Z(\eta)}{\delta \eta(t,x)},\phi(t,x)\rangle,
\end{aligned}
$$
this yields formally
$$
\begin{aligned}
\frac{\delta Z(\eta)}{\delta \eta(t,x)}&=\int d_F u\, i\hbar^{-1}u(t,x)e^{i\hbar^{-1}S(u)}e^{i\hbar^{-1}\langle u,\eta\rangle},\\
\frac{\delta^3 Z(\eta)}{\delta \eta^3(t,x)}&=\int d_F u\, (i\hbar^{-1} u(t,x))^3 e^{i\hbar^{-1}S(u)}e^{i\hbar^{-1}\langle u,\eta\rangle},
\end{aligned}
$$
and
$$
-i\hbar\square\frac{\delta Z(\eta)}{\delta \eta(t,x)}+i\lambda \hbar^3\frac{\delta^3 Z(\eta)}{\delta \eta^3(t,x)}
=\int d_F u\, ( \square u(t,x)+\lambda u^3(t,x))\, e^{i\hbar^{-1}S(u)}e^{i\hbar^{-1}\langle u,\eta\rangle}.
$$
Combining this with \eqref{NLW}, we have
$$
-i\hbar\square\frac{\delta Z(\eta)}{\delta \eta(t,x)}+i\lambda \hbar^3\frac{\delta^3 Z(\eta)}{\delta \eta^3(t,x)}
=-\int d_F u\, \frac{\delta S(u)}{\delta u(t,x)}e^{i\hbar^{-1}S(u)}\,e^{i\hbar^{-1}\langle u,\eta \rangle}.
$$
Remarking
$$
\frac{\delta e^{i\hbar^{-1}S(u)}}{\delta u(t,x)}=i\hbar^{-1}\frac{\delta S(u)}{\delta u(t,x)}e^{i\hbar^{-1}S(u)},
$$
and assuming that integration by parts was allowed, we get formally
\begin{equation}
-i\hbar\square\frac{\delta Z(\eta)}{\delta \eta(t,x)}+i\lambda \hbar^3\frac{\delta^3 Z(\eta)}{\delta \eta^3(t,x)}=\eta(t,x) Z(\eta)
\with Z(0)=1.
\label{phi4SD}
\end{equation}

\subsection{A Schwinger-Dyson equation corresponding to the anharmonic operator}
Since it may be thought that the quantum field theory has not yet mathematically settled, we need to check above procedure
comparing with the Schr\"odinger equation which seemingly corresponds to QFT with $d=0$.

Take an action integral corresponding to the anharmonic operator
\begin{equation}
S(q)=\int_{\euc} dt L(q,\dot{q}) \where L(q,\dot{q})=\frac{1}{2}\dot{q}^2(t)-V(q(t)) \with V(q)=\frac{\omega^2}{2}q^2-\frac{\lambda}{4} q^4.
\end{equation}
Therefore, the critical point  of $S$ or $L(q,\dot{q})$ is given by
\begin{equation}
0=\frac{\delta S(q)}{\delta q(t)}=-\ddot q(t)-\omega^2 q(t)-\lambda q^3(t)\quad\mbox{or}\quad
\dt \bigg(\frac{\partial L(q,\dot{q})}{\dot{q}}\bigg)-\frac{\partial V(q)}{\partial q}=0.
\label{CManh}
\end{equation}
From Lagrangian $L(q,\dot{q})$, by Legendre transformation, we have Hamiltonian $H(q,p)$ as
$$
H(q,p)=\frac{1}{2}p^2+V(q).
$$
Corresponding to this \eqref{CManh}, we may associate its quantum operator and Schr\"odinger equation as
\begin{equation}
i\hbar u_t=\hat{H}u \with \hat{H}(q,\partial_q)=-\frac{\hbar^2}{2}\partial_{q}^2-\lambda q^3.
\end{equation}

Analogously as above, we need to consider the characteristic functional to the measure $\displaystyle{d_F q\,e^{i\hbar^{-1}S(q)}}$:
$$
Z(\eta)=\int d_F q \,e^{i\hbar^{-1}S(q)}e^{i\hbar^{-1}(q,\eta)} \with \langle q,\eta\rangle=\int_{\euc}dt\, q(t)\eta(t).
$$
Therefore, the procedure of time-slicing method seems to regard the ``probability'' of $d_F q \,e^{i\hbar^{-1}S(q)}$
restricting to a subset
$$
{\mathcal{M}}([0,t]:\unbq, \bar{q})=\{q(\cdots)\;|\; q(0)=\unbq, q(t)=\bar{q}\}.
$$
\begin{problem}
Formulate above vague expression as mathematical statement and prove it!
Moreover, connect these with obtained results such as Caliceti \cite{cal99}, 
Caliceti, Graffi and Maioli \cite{CGM80}, 
Bender-Wu \cite{BW69} 
and Graffi, Grecchi and Simon \cite{GGS70}. 
\end{problem}

\subsection{A Schwinger-Dyson equation corresponding to the coupled Maxwell-Dirac equation}
As is mentioned in Dowling \cite{dow89} 
that Quantum electrodynamics(=QED) is a theory which describes how point, charged particles interact with light. Some of the theoretical predictions of QED agree with experiment to within one part in $10^{12}$ -- making it the most accurate physical theory ever invented.

For reader's sake, I explain the coupled Maxwell-Dirac equation and its quantized FDE following \cite{IZ79}. 
\paragraph{\bf The coupled Maxwell-Dirac equation}
The interaction of an electron with its self-induced electromagnetic field is governed by 
\begin{equation}
\left\{
\begin{aligned}
&(-i\gamma^{\mu}\partial_{\mu}+m)\psi=eA_{\mu}\gamma^{\mu}\psi,\\
&\square A_{\mu}=e\bar{\psi}\gamma_{\mu}\psi, 
\end{aligned}
\right.
\label{cMD}
\end{equation}
with the initial condition given by
\begin{equation}
\psi(0,x)=\psi_0(x), \; A_{\mu}(0,x)=a_{\mu}(x), \;\partial_t A_{\mu}(0,x)=b_{\mu}(x).
\label{ivcMD}
\end{equation}
Here, $e$ the electric charge, $m\ge0$ the mass of electron, $\psi={}^t(\psi_1,{\cdots}\psi_4)\in{\mathbb{C}}^4$ the $4$-spinor (the Dirac wave-function) and $\bar{\psi}$ the Dirac conjugate. $\partial_{\mu}={\partial}/{\partial x_{\mu}}$, i.e. $\partial_0=\partial_t$,
Minkowski metric $(g^{\mu\nu})={\rm diag}(1,-1, -1, -1)$,
$$
\gamma^{\mu}\gamma^{\nu}+\gamma^{\nu}\gamma^{\mu}=2g^{\mu\nu}{\mathbb{I}}_4, \; (\gamma^0)^*=\gamma^0, \; (\gamma^j)^*=-\gamma^j
$$
$\psi^*$ the complex conjugate transpose of $\psi$, $\bar{\psi}=\psi^*\gamma^0$.
$$
\gamma^0=\begin{pmatrix}
{\mathbb{I}}_2&0\\
0&-{\mathbb{I}}_2
\end{pmatrix}={\pmb{\sigma}}^3\otimes{\mathbb{I}}_2, \;
\pmb{\gamma}=i{\pmb{\sigma}}^2\otimes{\pmb{\sigma}}=\begin{pmatrix}
0&{\pmb{\sigma}}\\
-{\pmb{\sigma}}&0
\end{pmatrix},
$$
$$
\sigma^1=\begin{pmatrix}
0&1\\
1&0  
\end{pmatrix},\;
\sigma^2=\begin{pmatrix}
0&-i\\
i&0
\end{pmatrix},\;
\sigma^3=\begin{pmatrix}
1&0\\
0&-1
\end{pmatrix}.
$$
Or more precisely,
$$
\gamma^0=\begin{pmatrix}
1&0&0&0\\
0&1&0&0\\
0&0&-1&0\\
0&0&0&-1
\end{pmatrix},\;
\gamma^1=\begin{pmatrix}
0&0&0&1\\
0&0&1&0\\
0&-1&0&0\\
-1&0&0&0
\end{pmatrix},\;
\gamma^2=\begin{pmatrix}
0&0&0&-i\\
0&0&-i&0\\
0&i&0&0\\
i&0&0&0
\end{pmatrix},\;
\gamma^3=\begin{pmatrix}
0&0&1&0\\
0&0&0&-1\\
1&0&0&0\\
0&-1&0&0
\end{pmatrix}.
$$

\begin{problem}
Though there are many papers concern with the initial value problem of \eqref{cMD} with \eqref{ivcMD}
such as Gross \cite{gross66}], Chadam \cite{cha72}, Flato, Simon and Taflin \cite{FST87}, Bournaveas \cite{bou97}, Selberg and Tesfahun \cite{ST21} and Praselli \cite{pra05}, but I don't know whether there exists a global in time ``weak solution" of them, like such solutions for the Navier-Stokes equation.
\end{problem}

\paragraph{\bf A Schwinger-Dyson equation corresponding to \eqref{cMD}}
Above equation appears as the critical point of the following action functional: 
For each $(A,\psi,\bar{\psi})$, we define
\begin{equation}
{\mathcal{S}}(A,\psi,\bar{\psi})=\int_{{\mathbb{R}}^4}dxdt {\mathcal{L}}(A,\psi,\bar{\psi})
\end{equation}
with
\begin{equation}
\begin{aligned}
{\mathcal{L}}(A,\psi,\bar{\psi})&={\mathcal{L}}_0^\gamma+{\mathcal{L}}_0^{e^+, e^-}+{\mathcal{L}}_{int},\\
&=\frac{{\mathbb{E}}^2-{\mathbb{B}}^2}{2}+\frac{i}{2}\bar{\psi}\gamma^{\mu}\overset{\leftrightarrow}{\partial_{\mu}}\psi-m\bar{\psi}\psi-e\bar{\psi}\gamma^{\mu}\psi A_{\mu}.
\end{aligned}
\end{equation}
Here, $\mathbb{A}=(A^0,{\bf A}), \; \bf A=(A^1,A^2, A^3)$, $\psi={}^t(\psi_1,\psi_2,\psi_3,\psi_4)\in{\mathbb{C}}^4$,
$\bar{\psi}=(\bar{\psi}_1,\bar{\psi}_2,\bar{\psi}_3,\bar{\psi}_4)\in{\mathbb{C}}^4$, and
$$
u\overset{\leftrightarrow}{\partial}v=u(\partial v)-(\partial u)v,\quad
\mathbb{E}=-\nabla A^0-\frac{\partial {\bf A}}{\partial t},\quad    
\mathbb{B}={\rm{rot }}{\bf A}=\nabla\times{\bf A}=\begin{pmatrix}
\frac{\partial A^3}{\partial x_2}-\frac{\partial A^2}{\partial x_3}\\
\frac{\partial A^1}{\partial x_3}-\frac{\partial A^3}{\partial x_1}\\
\frac{\partial A^2}{\partial x_1}-\frac{\partial A^1}{\partial x_2}
\end{pmatrix}.
$$
From above, we get
\begin{equation}
\begin{aligned}
\frac{\delta \mathcal{S}(A,\psi,\bar{\psi})}{\delta A_{\mu}(t,x)}&=-\square A_{\mu}(t,x)-e\bar{\psi}(t,x)\gamma^{\mu}\psi(t,x), \\
\frac{\delta \mathcal{S}(A,\psi,\bar{\psi})}{\delta \psi(t,x)}&=(-i\gamma^{\mu}\partial_{\mu}-m)\bar{\psi}(t,x)-eA_{\mu}(t,x)\bar{\psi}(t,x)\gamma^{\mu},\\
\frac{\delta \mathcal{S}(A,\psi,\bar{\psi})}{\delta \bar\psi(t,x)}&=(i\gamma^{\mu}\partial_{\mu}-m)\psi(t,x)-eA_{\mu}(t,x)\gamma^{\mu}\psi(t,x).
\end{aligned}
\end{equation}

As explained before, we need to characterize the ``measure" $\displaystyle{D_F(A,\psi,\bar{\psi})e^{i\hbar^{-1}{\mathcal{S}}(A,\psi,\bar{\psi})}}$ from its characteristic functional:
$$
Z(J,\eta,\bar{\eta})=\int D_F(A,\psi,\bar{\psi})\,e^{i\hbar^{-1}{\mathcal{S}}(A,\psi,\bar{\psi})} e^{i\hbar^{-1}\langle(J,\eta,\bar{\eta}),(A,\psi,\bar{\psi})\rangle}
$$
where $\displaystyle{D_F(A,\psi,\bar{\psi})=d_FA \,d_F\psi \,d_F{\bar{\psi}}}$ and
$$
\langle(J,\eta,\bar{\eta}),(A,\psi,\bar{\psi})\rangle=\int dt\,dx (J_{\mu}(t,x)A^{\mu}(t,x)+\bar{\eta}(t,x)\psi(t,x)+\bar{\psi}(t,x)\eta(t,x)).
$$
Therefore, we have
$$
\begin{aligned}
\frac{\delta Z(J,\eta,\bar{\eta})}{\delta J_{\mu}(t,x)}&=i\hbar^{-1}\int D_F(A,\psi,\bar{\psi})\,A^{\mu}(t,x)e^{i\hbar^{-1}{\mathcal{S}}(A,\psi,\bar{\psi})}e^{i\hbar^{-1}\langle(J,\eta,\bar{\eta}),(A,\psi,\bar{\psi})\rangle}\\
\frac{\delta Z(J,\eta,\bar{\eta})}{\delta \eta(t,x)}&=i\hbar^{-1}\int D_F(A,\psi,\bar{\psi})\,\bar{\psi}(t,x)e^{i\hbar^{-1}{\mathcal{S}}(A,\psi,\bar{\psi})}e^{i\hbar^{-1}\langle(J,\eta,\bar{\eta}),(A,\psi,\bar{\psi})\rangle},\\
\frac{\delta Z(J,\eta,\bar{\eta})}{\delta  \bar{\eta}(t,x)}&=i\hbar^{-1}\int D_F(A,\psi,\bar{\psi})\,\psi(t,x)e^{i\hbar^{-1}{\mathcal{S}}(A,\psi,\bar{\psi})}e^{i\hbar^{-1}\langle(J,\eta,\bar{\eta}),(A,\psi,\bar{\psi})\rangle}.
\end{aligned}
$$
Moreover, we have very formally
$$
\begin{aligned}
\frac{\delta}{\delta \bar{\eta}(t,x)}\bigg(\gamma^{\mu} \frac{\delta Z(J,\eta,\bar{\eta})}{\delta {\eta}(t,x)}\bigg)
&=-\hbar^{-2}\int D_F(A,\psi,\bar{\psi})\,\bar{\psi}(t,x)\gamma^{\mu}{\psi}(t,x)\,e^{i\hbar^{-1}{\mathcal{S}}(A,\psi,\bar{\psi})}e^{i\hbar^{-1}\langle(J,\eta,\bar{\eta}),(A,\psi,\bar{\psi})\rangle},\\
\frac{\delta}{\delta J_{\mu}(t,x)}\bigg(\gamma^{\mu} \frac{\delta Z(J,\eta,\bar{\eta})}{\delta {\eta(t,x)}}\bigg)
&=-\hbar^{-2}\int D_F(A,\psi,\bar{\psi})\, A^{\mu}(t,x)\gamma^{\mu} \bar{\psi}(t,x)\,e^{i\hbar^{-1}{\mathcal{S}}(A,\psi,\bar{\psi})}e^{i\hbar^{-1}\langle(J,\eta,\bar{\eta}),(A,\psi,\bar{\psi})\rangle},\\
\frac{\delta}{\delta J_{\mu}(t,x)}\bigg(\gamma^{\mu} \frac{\delta Z(J,\eta,\bar{\eta})}{\delta \bar{\eta}(t,x)}\bigg)
&=-\hbar^{-2}\int D_F(A,\psi,\bar{\psi})\,A^{\mu}(t,x)\gamma^{\mu}{\psi}(t,x)
\,e^{i\hbar^{-1}{\mathcal{S}}(A,\psi,\bar{\psi})}e^{i\hbar^{-1}\langle(J,\eta,\bar{\eta}),(A,\psi,\bar{\psi})\rangle}.
\end{aligned}
$$
\begin{remark} 
To explain ``very formally", we take the first equation above as an example: Since
$$
\frac{d}{d\epsilon_2}\bigg[\frac{d}{d\epsilon_1}Z(J, \eta+\epsilon_2\phi_1,\bar{\eta}+\epsilon_2\bar{\phi}_2)\bigg|_{\epsilon_1=0}\bigg]\bigg|_{\epsilon_2=0}
=\iint dt dx dt' dx'
\frac{\delta}{\delta \bar{\eta}(t',x')}\frac{\delta Z(J,\eta,\bar{\eta})}{\delta {\eta}(t,x)}\phi_1(t,x)\bar{\phi}_2(t',x')
$$
We assume as if the following ``trace" exists
$$
\lim_{(t',x')\to (t,x)}\iint dt dx dt' dx' \frac{\delta}{\delta \bar{\eta}(t',x')}
\bigg[\frac{\delta Z(J,\eta,\bar{\eta})}{\delta {\eta}(t,x)}\phi_1(t,x)\bigg]\bar{\phi}_2(t',x')
=\int dt dx \frac{\delta}{\delta \bar{\eta}(t,x)}\frac{\delta Z(J,\eta,\bar{\eta})}{\delta {\eta}(t,x)}\phi_1(t,x)\bar{\phi}_2(t,x).
$$
or
$$
\lim_{(t',x')\to (t,x)}\frac{\delta}{\delta \bar{\eta}(t',x')}
\bigg[\frac{\delta Z(J,\eta,\bar{\eta})}{\delta {\eta}(t,x)}\phi_1(t,x)\bigg]\bar{\phi}_2(t',x')
=\frac{\delta}{\delta \bar{\eta}(t,x)}\frac{\delta Z(J,\eta,\bar{\eta})}{\delta {\eta}(t,x)}\phi_1(t,x)\bar{\phi}_2(t,x).
$$
\end{remark}

As before, we have formally
$$
\begin{aligned}
&i\hbar\square\frac{\delta Z(J,\eta,\bar{\eta})}{\delta J_{\mu}(t,x)}+e\hbar^2\frac{\delta}{\delta \bar{\eta}(t,x)}\bigg(\gamma^{\mu} \frac{\delta Z(J,\eta,\bar{\eta})}{\delta {\eta}(t,x)}\bigg)
=J_{\mu}(t,x)Z(J,\eta,\bar{\eta}),\\
&i\hbar(-i\gamma^{\mu}\partial_{\mu}-m)\frac{\delta Z(J,\eta,\bar{\eta})}{\delta \eta(t,x)}-e\hbar^2\frac{\delta}{\delta J_{\mu}(t,x)}\bigg(\gamma^{\mu} \frac{\delta Z(J,\eta,\bar{\eta})}{\delta {\eta(t,x)}}\bigg)=\eta(t,x)Z(J,\eta,\bar{\eta}),\\
\end{aligned}
$$   
$$
i\hbar(-i\gamma^{\mu}\partial_{\mu}+m)\frac{\delta Z(J,\eta,\bar{\eta})}{\delta \bar{\eta}(t,x)}-e\hbar^2\frac{\delta}{\delta J_{\mu}(t,x)}\bigg(\gamma^{\mu} \frac{\delta Z(J,\eta,\bar{\eta})}{\delta \bar{\eta}(t,x)}\bigg)=\bar{\eta}(t,x)Z(J,\eta,\bar{\eta}).
$$

\begin{remark}
From my knowledge, there exists no paper arguing the solvability of above FDE directly.
\end{remark}

\vspace{6mm}

\appendix

\section{Some questions on Euler or Navier-Stokes equations}
Though the following results are rather recent one which make me not only astonish, but also such impression is criticized because `` These results don't match physical reality'' by physicists and rather bypassed such as Sakajo \cite{sak23} 
{saying, Euler equation admits various unphysical weak solutions. But remember that writing down his equation, Maxwell recognized the electromagnetic wave which is not considered physical reality.}.
Even though, one may construct a weak solution with compact support in a torus dimension 2, a gravitational equation which is derived from classical idea

On the other hand, by the theory of general relativity, the following  Einstein's field equation
$$
G_{\mu \nu }+\Lambda g_{\mu \nu }=\kappa T_{\mu \nu }
$$
is derived where $G_{\mu \nu }=R_{\mu \nu }- \frac{1}{2}Rg_{\mu \nu }$ is the Einstein tensor, $g_{\mu \nu }$ is the metric tensor, 
$\Lambda$ is the cosmological constant and $\kappa$  is the Einstein gravitational constant.
Since from this equation, it is claimed the universe begins abruptly by Big Bang, I feel some analogy with above result.

We investigate the initial-boundary value problem
\begin{equation}\tag{NS}
u_t-\nu\Delta u+u\nabla u-\nabla p=f,\quad {\rm div}\, u=0, \quad u(0,x)=u_0(x),\quad  u|_{\partial\Omega}=0,
\label{NS}
\end{equation}
or
\begin{equation}\tag{E}
u_t+u\nabla u-\nabla p=0, \quad {\rm div}\, u=f, \quad u(0,x)=u_0(x), \quad u\cdot {\mathbf n}=0
\label{E}
\end{equation}
If we introduce the idea of weak solution for the Euler equation by integration by parts with suitable test functions, we have a result for Euler equation, see, DeLellis and Szekelyhidi \cite{DLS14, DLS15}. 

\begin{thm}[Theorem 1.7 of Isett and Oh \cite{IO14}]{\rm(Onsager's conjecture on manifolds, sharp version)}. 
Let $(M,g_{jk})$ be a compact Riemannian manifold and $I \subseteq \euc$ an open interval. Let $(u^{\ell}, p)$ be a weak solution to the Euler equations on $I\times M$ such that $u^{\ell}\in L^3_t(I;B^{1/3}_{3,c_0}(M))\cap C_t(I;L^2(M))$. 
Then, conservation of energy \eqref{E} holds.
\end{thm}

Analogously, we have

\begin{thm}[Theorem1.2 of Non-uniqueness of weak solutions for (NS) Buckmaster and Vicol \cite{BV18}] 
There exists $\beta>0$, such that nonnegative smooth function $e(t):[0,T]\to \euc_{\ge}$, there exists $v\in C_t^0([0,T];H_x^\beta({\mathbb{T}}^3))$ a weak solution of Navier-Stokes equation \eqref{NS}, such that $\int_{{\mathbb{T}}^3}dx|v(x,t)|^2=e(t)$ for all $t\in[0,T]$. Moreover, the associated vorticity $\nabla\times v$ lies in $C_t^0([0,T];L_x^1({\mathbb{T}}^3))$.
\end{thm}

\begin{thm}[${\mathbb{R}}^2$: Scheffer \cite{sche93}] 
There exists a weak solution, not identically zero, of \eqref{E} in ${\mathbb{R}}^2$, having compact support in $(x,t)\in {\mathbb{R}}^2\times(0,T)$. 
\end{thm}

\begin{thm}[${\mathbb{T}}^2$: Shnirelman \cite{shn97}] 
There exists a weak solution, not identically zero, of \eqref{E} in $\mathbb{T}^2$, having compact support in $t\in(0,T)$. 
\end{thm}

\begin{thm}[Theorem 1.3 of Buckmaster, DeLellis and L. Szekelyhidi \cite{BDLS15}] 
For any $\epsilon>$, there exists a non-trivial continuous weak solution $v:\mathbb{T}^3\times\euc \to \euc^3$ of \eqref{E}, with $v\in L^1(C^{1/3-\epsilon})$ with compact support in time.
\end{thm}
This stands for the collapse of the uniqueness, that is, if there exists very rapidly oscillating exterior force, then there exists a weak solution having space-time compact support.

\begin{thm}[\cite{sche93, shn97, BDLS15 }] 
There exists a weak solution $u(x,t)\in L^2(\euc^2\times\euc)$ such that $u(x,t)\equiv 0$ for $|x|^2+|t|^2>1$.
\end{thm}

Concerning this, the following paragraph suggests us we need to use new test functions\footnote{something like new tools to recognize new desease} to recognize anomalous phenomena.
\begin{quotation}[Shnirelman \cite{shn00}] 
The weak solution constructed by \cite{sche93, shn97} is not in fact a solution; very strong external forces are present, but they are infinitely-fast oscillating in space, and therefore are indistinguishable from zero in the sense of distributions. The smooth test-functions are not ``sensitive" enough to ``feel" these forces. This is the fault of sensors, not of forces.
\end{quotation}

\begin{thm}[Theorem1.1 of Yu \cite{yu18}] 
Let $\Omega$ be a bounded domain in $\euc^3$ with $C^2$-boundary $\partial\Omega$.
Let $u\in L^{\infty}(0,T;L^2(\Omega))\cap L^2(0,T; H^1(\Omega))$ be a weak solution of the Navier-Stokes equation for any smooth test function $\varphi\in C^{\infty}(\euc^+\times\Omega)$ with compact support, and ${\rm div}\,\varphi=0$.
In addition, if 
$$
u\in L^p(0,T;L^q(\Omega))\cap L^s(0,T; B_s^{\alpha,\infty}(\Omega))
$$
for any $\frac{1}{p}+\frac{1}{q}\leq \frac12,\;q\ge 4,\; s>2$ and for any $\frac12+\frac{1}{s}<\alpha<1$.
Then, for any $t\in[0,T]$,
$$
\int_{\Omega}dx |u(t,x)|^2+2\nu\int_0^t\int_{\Omega}dt dx|\nabla u|^2=\int_{\Omega}dx |u_0|^2.
$$
\end{thm}

\section{H. Lewy's operator by the method of characteristics}
Here, we quote the famous H. Lewy's counter example and its explanation by H\"ormander \cite{hor63}. 
\begin{thm}
There exists a function $f=f(t,x_1,x_2)\in C^{\infty}(\euc^3)$ such that the equation
\begin{equation}
\partial_tu+i\partial_{x_1}u-2i(t+ix_1)\partial_{x_2}u=f
\label{HL0}
\end{equation}
doesn't have any distributional solution $u$ in any open no-void subset of $\euc^3$.
\end{thm}

L. H\"ormander gives an resolution to this fact.
Let a linear PDE with smooth coefficients be given
$$
P(x,D_x)=\sum_{|\alpha|\le m} a_{\alpha}(x)D^{\alpha} \with a_{\alpha}(x)\in C ^{\infty}(\Omega:{\mathbb{C}}).
$$
Defining
\begin{equation}
P_m(x,\xi)=\sum_{|\alpha|= m} a_{\alpha}(x)\xi^{\alpha},\quad
\bar{P}_m(x,\xi)=\sum_{|\alpha|= m} \overline{a_{\alpha}}(x)\xi^{\alpha}
\label{6.1.1H}
\end{equation}
and
\begin{equation}
C_{2m-1}(x,\xi)=\sum_{j=1}^d i(P_m^{(j)}\bar{P}_{m,j}-P_{m,j}\bar{P}_m^{(j)})\where P_m^{(j)}=\frac{\partial P_m(x,\xi)}{\partial \xi_j}, P_{m,j}=\frac{\partial P_m(x,\xi)}{\partial x_j}.
\label{6.1.2H}
\end{equation}
\begin{thm}[Theorem 6.1.1 in \cite{hor63}] \label{Ho1}
Suppose that the differential equation
\begin{equation}
P(x,D_x)u=f
\label{6.1.0H}
\end{equation}
has a solution $u\in{\mathcal{D}}'(\Omega)$ for every $f\in C_0^{\infty}(\Omega)$.
Then, we have
\begin{equation}
C_{2m-1}(x,\xi)=0 \quad\mbox{if}\quad P_m(x,\xi)=0,\quad x\in\Omega,\quad \xi\in\euc^d.
\label{6.1.3H}
\end{equation}
\end{thm}

\begin{thm}[Theorem 6.1.2 in \cite{hor63}]
Suppose that the coefficients of the operator $P(x,D)$ of order $m$ are $C^{\infty}(\Omega)$ and that \eqref{6.1.3H} is not valid for any open non-void set $\omega\subset\Omega$ when $x$ is restricted to $\omega$. Then, there exists functions $f\in{\mathcal{S}}(\Omega)$ such that the equation \eqref{6.1.0H} doesn't have any solution $u\in{\mathcal{D}}'(\omega)$ for any open non-void set $\omega\subset\Omega$. The set of such functions $f$ is of second category.
\end{thm}

Taking $\alpha, \beta\in\euc$ such that $\beta(1+\alpha)\neq0$, we generalize slightly the operator in \eqref{HL0} as
$$
L_{\alpha, \beta}(t,x,\partial_t,\partial_x)=\partial_t+i\alpha\partial_{x_1}+i\beta(t+ix_1)\partial_{x_2}.
$$
Then, we have
$$
P_1((t,x),(\tau,\xi))=\tau+i\alpha\xi_1+i\beta(t+ix_1)\xi_2,\quad
\bar{P}_1((t,x),(\tau,\xi))=\tau-i\alpha\xi_1-i\beta(t-ix_1)\xi_2
$$
which implies
$$
C_1((t,x),(\tau,\xi))=-2\beta(1+\alpha)\xi_2
$$
and assuming $\beta(1+\alpha)\neq0$, we have the criterion of Theorem \ref{Ho1}:
$$
P_1((t,x),(\tau,\xi))=0 \quad\mbox{but}\quad C_1((t,x),(\tau,\xi))\neq 0 \quad\mbox{if}\quad \tau+\beta x_1=0 \et \beta t+\alpha\xi_1=0.
$$
Therefore,  H. Lewy example is reproduced by $(\alpha,\beta)=(1,-2)$.

In spite of above, we try to construct a solution of the following initial value problem:
\begin{equation}
L_{\alpha, \beta}(t,x,\partial_t,\partial_x)u(t,x)=f(t,x)\with  u(0,x)=\unbu(x).    
\label{HLmoc}
\end{equation}
Though the coefficients are complex valued, we may apply Theorem \ref{moc} formally.
The characteristic equation corresponding to \eqref{HLmoc} is
\begin{equation}
\begin{aligned}
&\frac{d x_1}{dt}=i\alpha \with x_1(0)=\underline{x}_1\\
&\frac{d x_2}{dt}=i\beta(t+ix_1) \with x_2(0)=\underline{x}_2.
\end{aligned}
\end{equation}
Therefore
\begin{equation}
\begin{aligned}
&x_1(t)=\underline{x}_1+i\alpha t,\\
&x_2(t)=\underline{x}_2-\beta\underline{x}_1t+i\beta(1-\alpha)\frac{t^2}{2}.
\end{aligned}
\end{equation}
Putting $x(t,\underline{x})=(x_1(t,\unbx), x_2(t,\unbx))$ and $F(t,\underline{x})=f(t, x(t,\underline{x}))$, we define
$$
U(t,\underline{x})=\unbu(x(t,\underline{x}))+\int_{0}^t ds\,F(s,\underline{x}).
$$
Denoting the inverse function of $\bar{x}=x(t,\underline{x})$ as $\underline{x}=y(t,\bar{x})$, more precisely,
\begin{equation}
\begin{aligned}
&\underline{x}_1=\bar{x}_1-i\alpha t=y_1(t,\barx),\\
&\underline{x}_2=\bar{x}_2+\beta \bar{x}_1 t-i\beta(1+\alpha)\frac{t^2}{2}=y_2(t,\barx),
\end{aligned}
\end{equation}
we define
\begin{equation}
u(t,\bar{x}) =U(t, x(t,\unbx))\big|_{\unbx=y(t,\bar{x})}.
\end{equation}
Finally, we have a solution of \eqref{HLmoc} explicitly if we may give the meaning to this expression: 
\begin{equation}
\begin{aligned}
u(t,\bar{x})&=\unbu(\bar{x}_1-i\alpha t,\bar{x}_2+\beta \bar{x}_1 t-i\beta(1+\alpha)\frac{t^2}{2})\\
&\qquad\qquad
+\int_{0}^t  ds f(s, x(s,\unbx))\bigg|_{\underline{x}=y(t,\bar{x})}.
\end{aligned}
\end{equation}

In fact, if the function $\unbu(x_1, x_2)$ is real analytic, then putting
$$
v(t,x)=\unbu({x}_1-i\alpha t,{x}_2+\beta  {x}_1 t-i\beta(1+\alpha)\frac{t^2}{2}),
$$
we have
$$
v_t+i\alpha v_{x_1}+i\beta(t+ix_1)v_{x_2}
=0  \with v(0,x)=\unbu(x).
$$
Moreover,
$$
\begin{aligned}
G(t,x)&=\int_0^t ds f(s,\unbx_1-i\alpha s,\unbx_2-\beta \unbx_1 s+i\beta(1+\alpha)\frac{s^2}{2})\bigg|_{\unbx=y(t,\barx), \barx=x}\\
&=\int_0^t ds f(s,x_1-i\alpha(t-s), x_2+\beta x_1(t-s)-i\alpha\beta\frac{(t-s)^2}{2}-i\beta\frac{t^2-s^2}{2})
\end{aligned}
$$
satisfies
$$
G_t+i\alpha G_{x_1}+i\beta(t+ix_1) G_{x_2}=f(t,x) \with G(0,x)=0.
$$
\begin{remark}
By Cauchy-Kowalevsky Theorem, we have a unique solution of \eqref{HLmoc} if coefficients are real analytic, therefore above arguments only imply the explicit representation of the solution.
\end{remark}

We quote here a result  mentioned to me by a letter from Takahiro Kawai on 01/Feb/1993:
\begin{claim} Take any functions $\phi, \psi$, for example, in $C^1(\euc)$.
Let a function $\unbu(x)=x_1\phi(x_2)+\psi(x_2)$ be given as initial value at time $t=0$ with $f=0$.
Assuming $\alpha=-1$, then \eqref{HL0} with $f=0$ and initial data $\unbu(x)$ has a unique solution
$$
u(t,x)=(x_1-i\alpha t)\phi(x_2+\beta x_1t)+\psi(x_2+\beta x_1t)
$$
of \eqref{HL0}.
\end{claim}
\par
{\it Proof}. Clear from above calculation because the function $x_1$ itself is real analytic.

\end{document}